# Resonant Interaction of Modulation-correlated Quantum Electron Wavepackets with Bound Electron States

Avraham Gover[1], Bin Zhang[1], Du Ran[1,2], Reuven Ianconescu[1,3], Aharon Friedman[4], Jacob Scheuer[1], Amnon Yariv[5]

[1] Department of Electrical Engineering Physical Electronics, Tel Aviv University, Ramat Aviv 69978, Israel

[2] School of Electronic Information Engineering, Yangtze Normal University, Chongqing 408100, China

[3] Shenkar College of Engineering and Design 12, Anna Frank St., Ramat Gan, Israel

[4] Ariel University, Ariel 40700, Israel

[5] California Institute of Technology (Caltech), Pasadena, California 91125, USA

*Free-Electron Bound-Electron Resonant Interaction (FEBERI) is the resonant inelastic interaction of periodically density-bunched free electrons with a quantum two level system. We present a comprehensive relativistic quantum mechanical theory for this interaction in a model in which the electrons are represented as quantum electron wavepackets (QEW). The analysis reveals the wave-particle duality nature of the QEW, delineating the point-particle-like and wave-like interaction regimes, and manifesting the physical reality of the wavefunction dimensions and its density modulation characteristics in interaction with matter. The analysis comprehends the case of laser-beam-modulated multiple QEWs that are modulation-phase correlated. Based on the Born interpretation of the electron wavefunction we predict quantum transitions enhancement proportional to the number of electrons squared, analogous to superradiance.*

We present here a comprehensive quantum model for the recently proposed new concept of Free-Electron-Bound-Electron Resonant Interaction (FEBERI) [1]. It has been asserted that in this process, pre-shaping of the quantum electron wavefunction of a free electron interacting with a bound electron, affects the probability of transition between the bound electron quantum levels, manifesting the reality of the Quantum Electron Wavepacket (QEW) in electron-matter interaction. In particular, a probability-density modulated QEW interacts resonantly with a Two-Level System (TLS) when its optical frequency modulation matches the TLS quantum energy level transitions:

$$n\hbar\omega_b = E_{2,1} \qquad (1)$$

where $\omega_b$ is the periodic temporal modulation frequency of the QEW density distribution as seen by a stationary observer, equal to the laser beam frequency that modulates it. $n\omega_b$ is an harmonic frequency of the QEW periodic bunching and $E_{2,1} = E_2 - E_1$ is the quantum energy gap of the TLS. This assertion has raised a debate [53, 54], but also a stream of numerous recently published papers relating to different aspects of this effect and its potential applications in electron microscopy, and in diagnostics and coherent control of q-bits and quantum emitters that are based on TLS [55-59].

Optical frequency modulation of single quantum electron wavepackets is a direct outcome of the process of Photon-Induced Near-Field Electron Microscopy (PINEM) [2- 12]. In this process, the energy spectrum of single Quantum Electron Wavepackets (QEW) is modulated at optical frequencies by interaction with a laser beam of frequency $\omega_b$, exhibiting discrete energy sidebands $\Delta E_n = n\hbar\omega_b$. The interaction is made possible by a multiphoton emission/absorption process in the near field of a nanostructure [8, 10], a foil [9,7] or a laser-beat (pondermotive potential) [11, 12]. It was also shown [8,9] that due to the nonlinear energy dispersion of electrons in free space drift, the discrete energy modulation of the QEW is converted into tight periodic density modulation (bunching) at atto-second short levels, corresponding to high spectral harmonics contents $\omega_n = n\omega_b$ in the expectation value of the QEW density:.

$$n(\mathbf{r},t) = \left\langle \left|\Psi(\mathbf{r},t)\right|^2 \right\rangle \tag{2}$$

The interpretation of the quantum electron wavefunction (QEW) has been a matter of debate since the inception of quantum theory [13, 14]. The accepted Born interpretation of the quantum electron wavefunction is that the expectation value (2) represents the probability density of finding the electron at point **r** at time t. The semiclassical interpretation of $-e\langle|\Psi(\mathbf{r},t)|^2\rangle$ as charge density may have limited validity in situations where it is possible to take an ensemble average over multiple electrons [15]. The reality of the QEW and the measurability of its dimensions, and the transition from the quantum wavefunction presentation to the classical point-particle theory (the wave-particle duality) were considered recently in the context of radiative interaction of single electron QEWs with light near polarizable structures, such as in Smith-Purcell radiation [16-18, 46]. Semiclassical analysis of such <u>stimulated</u> interaction (under external radiation field) reveals the transition from the quantum electron wavefunction interaction regime that is characterized by the multi-sidebands ($\Delta E_n = n\hbar\omega_b$) electron energy spectrum of PINEM [2-12] to the classical point-particle-like acceleration\deceleration regime. This transition takes place when

$$\Gamma = \omega\sigma_{et} = 2\pi\sigma_{ez}/\beta_0\lambda < 1, \tag{3}$$

where $\beta_0 = v_0/c$ is the centroid velocity of the QEW [16, 45]. Namely, the transition takes place when the wavepacket duration $\sigma_{et}$ or its length $\sigma_{ez}$ are short relative to the optical radiation period $2\pi/\omega$ or wavelength $\lambda$ respectively. This result, indicating the reality and measurability of the QEW dimensions in interaction with light, has been confirmed also by a QED theory analysis [19].

Furthermore, both semiclassical and QED model analyses [18,19] suggest that stimulated interaction of radiation with modulated QEWs are sensitive also to modulation features of the QEW. So, the density bunching of the QEW after a PINEM interaction and a subsequent free drift step, is measurable by interaction with a second synchronous laser beam. The physical reality of the periodic sculpting of the QEW in the time and space (propagation coordinate – z) dimensions, has been demonstrated recently experimentally by interaction with a second laser beam, phase-locked to the bunching frequency $\omega_b$ or its harmonic [9,10,20,21].

The physical reality of the wavepacket dimensions and its modulation features in the case of spontaneous radiative interaction in different schemes of QEW interactions with radiation is still being studied, and is a subject of controversies [22-24]. QED analyses of spontaneous Smith-Purcell radiative emission by a single QEW disaffirms dependence on the transverse dimensions of the QEW [24] and on its longitudinal dimension and modulation features [19]. However, the case may be different when multiple electrons are considered. In the classical point-particle regime, a multiple particles beam exhibits an effect of spontaneous superradiance (in the sense of Dicke [25]) when bunched within less than an optical wavelength [26]. Namely, they emit coherent radiation proportional to the square of the number of particles - $N^2$. One would expect that a bunch of identical QEWs that satisfy condition (3) would likewise exhibit similar $N^2$ scaling [18], and that it would turn into the N scaling dependence of "shot-noise radiation" when the condition is not satisfied. This establishes a dependence of the coherent superradiance emission on the QEWs dimension in the case of multiple QEWs. Furthermore, as reviewed in [27], periodically bunched point-particle electron beams emit superradiantly at the frequency and harmonics of their density modulation frequency with the same quadratic scaling. Based on the Born probability interpretation, one would expect that similar $N^2$ scaling of superradiant emission will take place also with a multi-QEWs beam when the density expectation value of their individual wavefunctions is modulated at optical frequency, under the condition that their modulation phases are correlated. Such a correlated QEWs beam can be generated in the PINEM interaction process if all QEWs are exposed at the PINEM interaction point to the near field of the same coherent laser beam as in [8]. In this case of multiple particles, a semiclassical model is valid [18] and predicts the same quadratic scaling of superradiant emission by bunched QEWs.

In the FEBERI concept of Ref. 1 the idea of the reality of the QEW modulation features is implemented in the case of interaction with matter. Ref. 1 used a simple semiclassical model in order to bring to light the feasibility of the FEBERI effect. In this model the modulated QEW probability density is taken to represent periodic space-charge distribution of the QEW. Such a semiclassical model is questionable in the case of single electrons, but is expected to have partial validity in appropriate limits of multiple correlated QEWs. Nevertheless, a more complete quantum mechanical theory analysis is called for, in order to affirm the feasibility of the effect and the limits of its validity. Furthermore, the interaction with multiple QEWs involves their entanglement with the bound electron (TLS) states and their entanglement with each other. In a future publication we expect to refer to this additional interesting feature of the multi-particle FEBER interaction. In the present publication we limit our scope to formulation of the FEBERI problem in terms of a complete quantum mechanical theory model, and to identifying the operating regimes where the semiclassical model [1] is valid.

## 1. Quantum Formulation of FEBERI

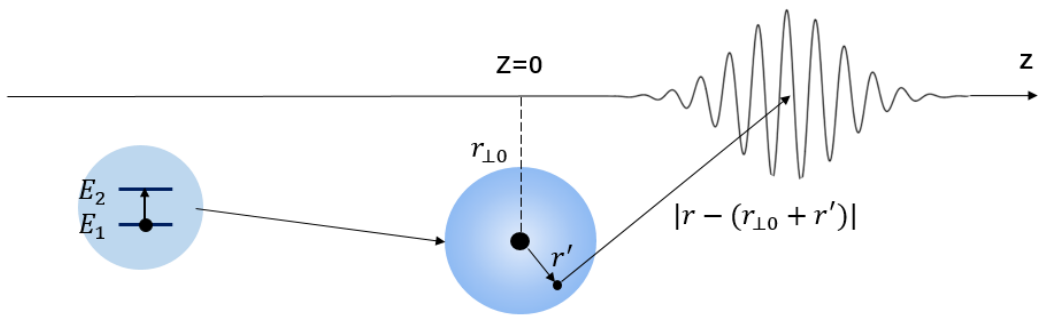

Fig. 1. A Two-Level System (TLS) quantum interaction model of a Quantum Electron Wavepacket interaction with a bound electron.

Our comprehensive quantum mechanical theory model is based on the setup shown in Fig. 1, showing a thin free electron QEW propagating in proximity to a TLS that is modeled as a Hydrogen-like atom. For simplicity we assume that the interaction of the free and bound electrons is only through Coulomb interaction. We start with a Schrodinger equation for the combined wavefunction of the free and bound electrons:

$$i\hbar \frac{\partial \Psi(\mathbf{r},\mathbf{r}',t)}{\partial t} = (H_0 + H_I)\Psi(\mathbf{r},\mathbf{r}',t) \tag{4}$$

$$H_0 = H_{0F} + H_{0B} \tag{5}$$

where $H_{0F}, H_{0B}$ are the Hamiltonians of the free and bound electrons respectively, and $H_I$ is the interaction Hamiltonian. In order to apply the analysis also to relativistic electrons, we use the "relativistic" Schrodinger equation Hamiltonian for a free electron of energy $\mathcal{E}_0 = \gamma_0 mc^2$ and momentum $\mathbf{p_0} = \gamma_0 m \mathbf{v_0}$:

$$H_{0F}(\mathbf{r}) = \mathcal{E}_0 + \mathbf{v}_0 \cdot (-i\hbar\nabla - \mathbf{p}_0) + \frac{1}{2\gamma_0^3 m}(-i\hbar\nabla - \mathbf{p}_0)^2 \tag{6}$$

This Hamiltonian was derived in Ref. [28] and the Supplementary of [16] by a second order iterative approximation of Klein-Gordon equation, and therefore does not include spin effects. It has been derived recently also directly from the Dirac equation [29] without the quadratic term that is needed here to account for electron wavepacket chirp and size expansion in free drift. It corresponds to second order expansion of the relativistic energy dispersion of a free electron:

$$E_p = \mathcal{E}_0 + \mathbf{v}_0 \cdot (\mathbf{p} - \mathbf{p}_0) + \frac{1}{2\gamma_0^3 m}(\mathbf{p} - \mathbf{p}_0)^2 \tag{7}$$

The eigenfunction solutions of the bound electron Hamiltonian $H_{0B}$ are modelled as the solutions of a two-level system (TLS):

$$H_{0B}\Psi_j(\mathbf{r}',t) = E_j \Psi_j(\mathbf{r}',t) \quad (j=1,2) \tag{8}$$

$$\Psi_j(\mathbf{r}',t) = \varphi_j(\mathbf{r}')e^{-iE_j t/\hbar} \tag{9}$$

$$\Psi_B(\mathbf{r}',t) = \sum_{j=1}^{2} C_j \Psi_j(\mathbf{r}',t) \tag{10}$$

The wavefunction solution of the free electron in zero order is taken to be a wavepacket:

$$\Psi_F^{(0)}(z,t) = \int \frac{dp}{\sqrt{2\pi\hbar}} c_p^{(0)} e^{-iE_p t/\hbar} e^{ipz/\hbar} \tag{11}$$

for a Gaussian wavepacket [16]:

$$c_p^{(0)}(t_0) = \frac{1}{(2\pi\sigma_{p0}^2)^{1/4}} e^{-(p-p_0)^2/4\tilde{\sigma}_p^2(t_0)} e^{i(p_0 L_D - E_0 t_D)/\hbar} \tag{12}$$

and in space coordinates:

$$\psi_F^{(0)}(z,t) = \frac{\sqrt{\sigma_{z_0}}}{\left(2\pi\tilde{\sigma}_z^4(t+t_D)\right)^{1/4}} \exp\left(-\frac{(z-v_0(t-t_0))^2}{4\tilde{\sigma}_z^2(t+t_D)}\right) e^{i(p_0(z+L_D)-E_0(t+t_D))/\hbar}, \qquad (13)$$

These equations for a freely drifting QEW, were derived in [16] for the Hamiltonian (6) that includes wavepacket chirp and expansion effects due to the dispersive second order term in (6). For simplicity we assume here that the QEW reaches the interaction point z=0 at time $t_0$ at its longitudinal waist, so that the complex parameters reduce at this point to be real: $\tilde{\sigma}_z(t) = \sigma_{z0}, \tilde{\sigma}_p = \sigma_{p0}$ and sigma$_{z0} = \hbar/2\sigma_{p0}$.

For a modulated wavepacket, with the same assumption [18]:

$$c_p^{(0)} = \left(2\pi\sigma_{p_0}^2\right)^{-1/4} \sum_{n=-\infty}^{\infty} J_n(2|g|) \exp\left(-\frac{(p-p_0-n\delta p)^2}{4\sigma_{p_0}^2}\right) e^{in\phi_b} e^{-i(p-p_0)^2 t_D/2m^*\hbar} e^{i(p_0 L_D - E_0 t_D)/\hbar} \qquad (14)$$

$$\psi_F^{(0)}(z,t) = \frac{e^{i(p_0 z - E_{p_0} t)/\hbar}}{\left(2\pi\sigma_{z_0}^2\right)^{1/4}} \sum_{n=-\infty}^{\infty} J_n(2|g|) \exp\left(-\frac{(z-v_0(t-t_0)-n\delta pt/2\gamma^3 m)^2}{4\sigma_{z_0}^2}\right) e^{i\left(\frac{n\omega_b}{v_0}\right)(z-v_0 t - n\delta pt/2\gamma^3 m)} \qquad (15)$$

where $\delta p = \hbar\omega_b/v_0$.

In a simplified model, the spin is neglected, and we assume that the free and bound electrons do not overlap spatially. Therefore, there are no exchange energy or spin–orbit interaction effects, and we can avoid the intricate second quantization of many body interaction theory [30]. We assume that the only interaction is Coulomb interaction, and in the near field, neglecting retardation [36] and with gauge **A** = 0, the interaction Hamiltonian is

$$H_I(\mathbf{r},\mathbf{r'}) = \frac{e^2}{4\pi\varepsilon_0} \frac{\gamma}{\left[(\gamma z - z')^2 + (\mathbf{r}_\perp - \mathbf{r'}_\perp)^2\right]^{1/2}} \qquad (16)$$

$$H_I(\mathbf{r},\mathbf{r'}) \simeq \frac{e^2}{4\pi\varepsilon_0} \left[\frac{1}{(\gamma^2 z^2 + r_{\perp 0}^2)^{1/2}} + \frac{\mathbf{r'} \cdot (\mathbf{e}_z \gamma z - \mathbf{e}_r \mathbf{r}_{\perp 0})}{(\gamma^2 z^2 + r_{\perp 0}^2)^{3/2}}\right] \qquad (17)$$

Here we used Feynman's expression for the Coulomb potential [31] in order to keep the analysis valid in the relativistic regime. A more accurate form would be to use the Darwin potential for relativistic Coulomb interaction between moving charged particle [33,34]. We believe, however that for the parameters of the cases delineated here, the corrections due to this model are negligible.

In the interaction process, the expansion coefficients of the QEW $c_p^{(0)}$ (12) or (14) are entangled with the coefficients of the bound electron $C_j$ (10) and the combined wavefunction is:

$$\Psi(\mathbf{r},\mathbf{r'},t) = \sum_{j=1}^{2} \int_p dp\, c_{j,p}(t) \varphi_j(\mathbf{r'}) e^{-iE_j t/\hbar} e^{-iE_p t/\hbar} e^{ipz/\hbar} \qquad (18)$$

and after substitution in (4):

$$i\hbar \frac{\partial \Psi}{\partial t} = i\hbar \sum_{j=1}^{2} \int_p dp \left[ \dot{c}_{j,p}(t) - \frac{E_j + E_p}{\hbar} c_{j,p}(t) \right] \varphi_j(\mathbf{r}') e^{-iE_j t/\hbar} e^{-iE_p t/\hbar} e^{ipz/\hbar} =$$

$$= \left( H_{0B} + H_{0F} + H_I \right) \sum_{j=1}^{2} \int_p dp\, c_{j,p}(t) \cdot \varphi_j(\mathbf{r}') \cdot e^{-iE_j t/\hbar} e^{-iE_p t/\hbar} e^{ipz/\hbar}$$

(19)

After cancelling out the no-interaction terms, we are left with:

$$i\hbar \sum_{j=1}^{2} \int_p dp\, \dot{c}_{j,p}(t) \varphi_j(r')\, e^{-\frac{iE_j t}{\hbar}} e^{-\frac{i(E_p t - pz)}{\hbar}} = H_I(r, r') \sum_{j=1}^{2} \int_p dp\, c_{j,p}(t)\, \varphi_j(r')\, e^{-\frac{iE_j t}{\hbar}} e^{-\frac{i(E_p t - pz)}{\hbar}} \quad (20)$$

We multiply by $\varphi_i^*(\mathbf{r}')$ and integrate over space. Using the ortho-normality relation

$$\int \varphi_i(\mathbf{r}') \varphi_j^*(\mathbf{r}') d^3r' = \delta_{i,j}:$$

$$i\hbar \int_p dp\, \dot{c}_{i,p}(t) e^{-iE_i t/\hbar} e^{-iE_p t/\hbar} e^{ipz/\hbar} =$$

$$e^{-iE_i t/\hbar} \int dp\, c_{i,p}(t) \langle i | H_I(\mathbf{r}, \mathbf{r}') | i \rangle e^{-iE_p t/\hbar} e^{ipz/\hbar} + e^{-iE_j t/\hbar} \int dp\, c_{j \neq i,p}(t) \langle i | H_I(\mathbf{r}, \mathbf{r}') | j \rangle e^{-iE_p t/\hbar} e^{ipz/\hbar}$$

(21)

where

$$M_{i,j}(\mathbf{r}) = \langle i | H_I(\mathbf{r}, \mathbf{r}') | j \rangle \equiv \int d^3r'\, \varphi_i^*(\mathbf{r}') H_I(\mathbf{r}, \mathbf{r}') \varphi_j(\mathbf{r}') \quad (22)$$

For simplicity we redefine the self-interaction terms, so that $\langle i | H_I(\mathbf{r}, \mathbf{r}') | i \rangle$, then:

$$i\hbar \int_p dp\, \dot{c}_{i,p}(t) e^{-iE_i t/\hbar} e^{-iE_p t/\hbar} e^{ipz/\hbar} = e^{-iE_j t/\hbar} \int dp\, c_{j \neq i,p}(t) M_{i,j}(\mathbf{r}) e^{-iE_p t/\hbar} e^{ipz/\hbar} \quad (23)$$

This is an integro-differential equation that needs to be solved as a function of time for the initial condition $c_{j,p}(t_0^-) = C_j^{(0)}(t_0^-) c_p^{(0)}$

If $|\mathbf{r}'| \ll |\mathbf{r} - \mathbf{r}'| \approx (r_{\perp 0}^2 + \gamma^2 z^2)^{1/2}$ then the integration over $\mathbf{r}'$ can be carried out independently of $\mathbf{r}$:

$$M_{i,j}(\mathbf{r}_{\perp 0}, \mathbf{r}) = \int \varphi_i^*(\mathbf{r}') H_I(\mathbf{r}, \mathbf{r}') \varphi_j(\mathbf{r}') d^3r' \quad (24)$$

For the interaction (17):

$$M_{i,j} = \frac{e^2}{4\pi\varepsilon_0} \frac{\mathbf{r}_{i,j} \cdot (\mathbf{e}_z \gamma z - \mathbf{e}_r r_{\perp 0})}{(\gamma^2 z^2 + r_{\perp 0}^2)^{3/2}} \quad (25)$$

where

$$-e\mathbf{r}_{2,1} \equiv -e \int \varphi_2^*(\mathbf{r}') \mathbf{r}' \varphi_1(\mathbf{r}') d^3r' \quad (26)$$

is the dipole transition matrix element $\boldsymbol{\mu}_{2,1} = -e\mathbf{r}_{2,1}$.

## 2. Projection to Momentum space

We project the integro-differential equation (23) onto momentum space by multiplying with $e^{ip'z/\hbar}$ and integrating over z. With $\int dz\, e^{i(p'-p)z/\hbar} = 2\pi\hbar\delta(p'-p)$, we get:

$$2\pi i\hbar^2 \int_p dp\, \dot{c}_{i,p}(t) e^{-iE_p t/\hbar} \delta(p'-p) = e^{-i(E_j-E_i)t/\hbar} \int_p dp\, c_{j,p}(t) \int_{-\infty}^{\infty} dz\, M_{i,j}(\mathbf{r}) e^{i(p-p')z/\hbar} e^{-iE_p t/\hbar} \qquad (27)$$

$$\dot{c}_{i,p'}(t) = \frac{1}{2\pi i\hbar^2} \int dp\, \tilde{M}_{i,j}(p'-p) c_{j,p}(t) e^{-i(E_p-E_{p'}-E_{i,j})t/\hbar} \qquad (28)$$

where

$$E_{i,j} = E_i - E_j \qquad (29)$$

$$\tilde{M}_{i,j}(p'-p) = \int_{-\infty}^{\infty} dz\, M_{i,j}(\mathbf{r}) e^{i(p-p')z/\hbar} \qquad (30)$$

In Append A we present the explicit expressions of $\tilde{M}_{i,j}(p)$ for the dipole matrix elements (25) of the longitudinally and transversely aligned dipoles. This function is related to the momentum state matrix element of the interaction Hamiltonian (25) by:

$$\tilde{M}_{i,j}(p'-p) = 2\pi\hbar \langle p' | M_{i,j}(\mathbf{r}) | p \rangle = 2\pi\hbar \langle p', i | H_I(\mathbf{r}, \mathbf{r}') | p, j \rangle$$

The differential equation (28) describes the dynamic evolution of the entangled free electron and bound electron in momentum space. Before the start interaction time $t_0^-$, the free and bound electrons are not entangled:

$$c_{j,p}(t) = C_j^{(0)}(t_0^-) c_p^{(0)} \qquad (31)$$

Equation (28) can be solved by an iterative process, in which we assume to first order that the free and bound electrons are disentangled and not evolving in time during an effective interaction time $t_{int}$:

$$c_{j,p}(t) \approx C_j^{(0)}(t_0) c_p^{(0)} \qquad (32)$$

Substituting this time-independent amplitude in the RHS of (28), it is possible to integrate over time, and get a factor:

$$\int_{t_0^-}^{t_0^+} dt\, e^{-i(E_p-E_{p'}-E_{i,j})t/\hbar} = e^{-i(E_p-E_{p'}-E_{i,j})t_0/\hbar} \frac{(t_0^+ - t_0^-)}{\hbar} \text{sinc}\left[(E_p - E_{p'} - E_{i,j})(t_0^+ - t_0^-)/2\hbar\right] =$$
$$e^{-i(E_p-E_{p'}-E_{i,j})t_0/\hbar} \frac{2t_{int}}{\hbar} \text{sinc}\left[(E_p - E_{p'} - E_{i,j}) t_{int}/\hbar\right] \to 2\pi\hbar\delta(E_p - E_{p'} - E_{i,j}) \qquad (33)$$

where the last limit is taken for an infinite interaction time. This dictates a conservation of energy transfer condition:

$$E_{p'} - E_p = -E_{i,j} \qquad (34)$$

This condition, used in the dispersion relation (7) determines the recoil momentum of the QEW during the interaction:

$$p_{rec} = p' - p = -E_{i,j}/v_0 \tag{35}$$

The momentum recoil $p_{rec}$ is defined here so that if the transition is from the lower level j = 1 to the upper level i = 2, the momentum recoil is negative, and vice versa. Here we used only the first order expansion of the dispersion equation (7) (the dispersive second order term would introduce a small interaction quantum recoil correction, differentiating up and down transitions [16,28] that can be neglected in the present context). Thus integration of (28) using (31),(32) results in:

$$c_{i,p'}(t_0^+) = C_i^{(0)}(t_0^-)c_{p'}^{(0)} + \Delta c_{i,p'} \tag{36}$$

$$\Delta c_{i,p'} = \frac{1}{i\hbar v_0}\tilde{M}_{i,j}(p_{rec})C_j^{(0)}(t_0)c_{p'-p_{rec}}^{(0)} \tag{37}$$

Neglecting dynamics of the TLS during the interaction, the transition probability is:

$$P_i(t_0^+) = \int_{p'}\left|c_{i,p'}(t_0^+)\right|^2 dp' = \int_{p'}\left|C_i(t_0)c_{p'}^{(0)}(t_0) + \Delta c_{i,p'}((t_0^+))\right|^2 dp' = \tag{38}$$
$$P_i^{(0)} + \Delta P_i^{(1)} + \Delta P_i^{(2)}$$

where

$$P_i^{(0)} = \left|C_i(t_0)\right|^2 \int_p \left|c_{p'}^{(0)}(t_0)\right|^2 dp' = \left|C_i(t_0)\right|^2 \tag{39}$$

is the initial occupation probability of level i, and the incremental probabilities are:

$$\Delta P_i^{(1)} = 2\mathrm{Re}\left[C_i^{(0)*}(t_0)\int_{-\infty}^{\infty}dp'c_{p'}^{(0)*}(t_0)\Delta c_{i,p'}\right] \tag{40}$$

$$\Delta P_i^{(2)} = \int_{-\infty}^{\infty}dp'\left|\Delta c_{i,p'}\right|^2 \tag{41}$$

Substituting (37) in (41), and integrating over momentum, using $\int dp\left|c_{p-p_{rec}}^{(0)}\right|^2 = 1$, one obtains:

$$\Delta P_i^{(2)}(t_0^+) = \int dp'\left|c_{i,p'}^{(0)}\right|^2 = \frac{1}{\hbar^2 v_0^2}\left|\tilde{M}_{i,j}(p_{rec})\right|^2\left|C_j^{(0)}(t_0)\right|^2 \tag{42}$$

In the case of excitation of a TLS from ground level: $C_2^{(0)}(t_0^-) = 0, C_1^{(0)}(t_0^-) = 1$, the excitation probability of the TLS is given by:

$$P_2(t_0^+) = \Delta P_2^{(2)} = \int dp'\left|c_{2,p'}^{(0)}\right|^2 = \frac{1}{\hbar^2 v_0^2}\left|\tilde{M}_{2,1}(p_{rec})\right|^2 \tag{43}$$

Evidently the excitation probability in this case is independent of the QEW shape or dimensions. However, considering the case where the TLS is in a superposition state at the interaction time, the first order incremental probability term (40) may be dominant. Substituting (37) in (40) and integrating over momentum results in:

$$\Delta P_i^{(1)} = \frac{2}{\hbar v_0} \text{Re}\left[C_i^{(0)*}(t_0) C_j^{(0)} (\tilde{M}_{ij}(p_{rec})/i) I(p_{rec})\right] \tag{44}$$

$$I(p_{rec}) = \int c_p^{(0)*} c_{p-p_{rec}}^{(0)} \, dp \tag{45}$$

This integral is evaluated in Append. B for a Gaussian distribution of the QEW:

$$I(p_{rec}) = e^{-\frac{1}{2}(p_{rec}/2\sigma_{po})^2} e^{-i\omega_{i,j} t_0} \tag{46}$$

Substituting $p_{rec} = \hbar\omega_{1,2}/v$, $\sigma_{z0} = \hbar/2\sigma_{p0}$, $\sigma_{t0} = \sigma_{z0}/v$, one gets:

$$\Delta P_i^{(1)} = \frac{2}{\hbar v_0} \text{Re}\left[C_i^{(0)*}(t_0) C_j^{(0)}(t_0) e^{-i\omega_{i,j} t_0} (\tilde{M}_{i,j}(p_{rec})/i)\right] e^{-\Gamma^2/2}$$
$$= \frac{2}{\hbar v_0} \left|\tilde{M}_{i,j}(p_{rec})\right| \left|C_i^{(0)*}(t_0) C_j^{(0)}(t_0)\right| e^{-\Gamma^2/2} \sin\varsigma \tag{47}$$

where we substituted $C_i^{(0)*}(t_0) C_j^{(0)}(t_0) = \left|C_i^{(0)*}(t_0) C_j^{(0)}(t_0)\right| e^{i\varphi}$ in terms of the dipole moment excitation amplitude and phase of the state in the TLS Bloch sphere presentation, defined $\varsigma = \omega_{i,j} t_0 - \varphi$ the phase of the QEW arrival time relative to the TLS dipole moment excitation phase, and

$$\Gamma = \frac{p_{rec}}{2\sigma_{p0}} = \frac{\hbar\omega_{1,2}}{2v\sigma_{p0}} = \omega_{1,2}\sigma_{t0} = 2\pi\frac{\sigma_{z0}}{\beta\lambda_{1,2}} \tag{48}$$

This means that the incremental probability is dependent on the wavepacket dimensions if the phase of the superposition state of the TLS is pre-determined, and it vanishes for a long wavepacket. Instructively, Eq. 47 and the decay constant (48) are analogous to the decay of QEW acceleration in the quantum limit, and the transition from classical point-particle to quantum regime PINEM interaction in stimulated radiative interaction of a finite size QEW (3) [16].

We note that the assumption that $c_{jp}(t)$ is disentangled and constant during the interaction time $t_{int}$, as assumed in Eq. 32, may have partial validity, and the transition of the sinc function to delta function in (33) is questionable when the interaction time $t_{int}$ is short:

$$E_{2,1} t_{int} < \hbar/2 \tag{49}$$

Or, with the assumption that $t_{int}$ is the longer of the interaction transit time $t_r = r_\perp/\gamma v_0$ and the wavepacket duration $\sigma_{et} = \sigma_{z0}/v_0$:

$$t_r, \sigma_{et} < t_{int} < \frac{\hbar}{E_{2,1}} = \frac{1}{\omega_{2,1}} \tag{50}$$

Note that the limit $\sigma_{et} < 1/\omega_{2,1}$ is exactly the near-point-particle limit (3), discussed in the introduction. Also, instructively, when $\omega_{2,1}\sigma_{et} < 1/2$, then $\sigma_{ep} > E_{2,1}/v_0$, namely, the momentum spread is larger than the recoil (35).

In the next section we present an alternative approximate solution of the Schrodinger equation that contrary to the approximation of stationary TLS in (32), takes into consideration the dynamics of the TLS transition, and may better describe the short interaction regime (50).

### 3. Probabilistic Model for FEBERI Interaction

The zero order iteration of the momentum projection equation (28) with the substitution of (32) helps to describe the modification of the QEW distribution due to the FEBERI interaction, and infers to the dynamics of the TLS only through the final conservation of energy condition. In the near-point-particle QEW limits (50) this approximation does not describe the TLS dynamics. Rather than continuing this iterative approach in the momentum space, and solving (28) without taking the delta function limit in (33), we present in this section an alternative approach, going back to the source equation (23) and solving it directly, with a similar first order iteration approximation, by substituting on the RHS of (23):

$$c_{j,p}(t) \approx C_j^{(0)}(t) c_p^{(0)} \tag{51}$$

which is the same as in the previous assumption (32), but here allowing development in time of the TLS and neglect the recoil dynamics of the QEW. After multiplying (23) by the complex conjugate of the free electron wavefunction (11) and integrating over space: $\int d^3r\, \psi_F^{(0)*}(r,t)$, one obtains:

$$\frac{i}{2\pi}\int_p dp' \int_p \dot{c}_{i,p}(t) c_{p'}^{(0)*} e^{i(E_{p'}-E_p)t/\hbar} \int_z dz\, e^{i(p-p')z/\hbar} = C_j^{(0)}(t) e^{i(E_i-E_j)t/\hbar} \int d^3r M_{i,j}(\mathbf{r}) \left|\psi_F^{(0)}(\mathbf{r},t)\right|^2 \tag{52}$$

With $\int dz\, e^{i(p'-p)z/\hbar} = 2\pi\hbar\delta(p'-p)$:

$$2\pi i\hbar \int_p dp'\, \dot{c}_{i,p}(t) c_p^{(0)*} = C_j^{(0)}(t) e^{i(E_i-E_j)t/\hbar} \int d^3r M_{i,j}(\mathbf{r}) \left|\psi_F^{(0)}(\mathbf{r},t)\right|^2 \tag{53}$$

This presentation is reminiscent of point-particle interaction with the Born quantum wavefunction probability $\left|\Psi_F^{(0)}(\mathbf{r},t)\right|^2$ of electron arrival time t at z=0.

It should be stressed that $\left|\Psi_F^{(0)}(\mathbf{r},t)\right|^2$ is not well determined for a single electron. We assume that it is possible to solve (53) with substitution of its expectation value - $\langle\left|\Psi_F^{(0)}(\mathbf{r},t)\right|^2\rangle$, and the solution will then represent the result of interaction with an ensemble of identical QEWs.

The probability distribution of a single electron QEW of narrow width is:

$$\left\langle \left|\Psi_F^{(0)}(\mathbf{r},t)\right|^2 \right\rangle = \delta(\mathbf{r}_\perp) f_{ez}(z - v_0(t-t_0)) = \delta(\mathbf{r}_\perp) f_{et}(t - t_0 - z/v_0)/v_0 \tag{54}$$

where $f_{et}$ is normalized over time. Then:

$$i\hbar \int_p dp' \dot{c}_{i,p}(t) c_p^{(0)*} = C_j^{(0)}(t) e^{i\omega_{i,j}t} f(t-t_0) \tag{55}$$

$$f(t-t_0) = \frac{1}{v_0} \int dz M_{i,j}(z) f_{et}(t-t_0 - z/v) \tag{56}$$

where $M_{i,j}(z)$ is given in (25).

With approximation (51), neglecting now the dynamics of the QEW around the interaction time $t_0$, we can turn (55) into coupled differential equations for the TLS (i, j = 1,2):

$$\dot{C}_i(t) = \frac{1}{i\hbar} C_j(t) e^{-i\omega_{i,j}t} f(t-t_0) \tag{57}$$

and after integration,

$$C_i(t_0^+) = C_i(t_0^-) + \frac{1}{i\hbar} \int_{t_0^-}^{t_0^+} dt\, C_j(t) e^{-i\omega_{i,j}t} f(t-t_0). \tag{58}$$

The weighed interaction probability $f(t-t_0)$ (56) depends on the z dependence of both the interaction Hamiltonian (25) and the quantum probability (2), which for a Gaussian or modulated Gaussian QEW, can be calculated from the wavepacket functions (13) or (15).

For a single Gaussian wavepacket (13) at its longitudinal waist ($\tilde{\sigma}_z(t) = \sigma_{z0} = v_0 \sigma_{et}$):

$$f_{et}(t-t_0 - z/v) = \frac{1}{\sqrt{2\pi}\sigma_{et}} e^{-(t-t_0-z/v)^2/2\sigma_{et}^2} \tag{59}$$

Normalizing time to the transit time parameter $\bar{t} = t/t_r$, and defining $\bar{t}' = z/v_0 t_r$, the weighed interaction probability function (56) can be recast into a convolution relation with the ratio $\bar{\sigma}_{et} = \sigma_{et}/t_r$ between the wavepacket duration and the transit time as a parameter:

$$f_\parallel(t-t_0) = K_\parallel \int_{-\infty}^{\infty} d\bar{t}' \frac{\bar{t}'}{(\bar{t}'^2+1)^{3/2}} \frac{1}{\sqrt{2\pi}\bar{\sigma}_{et}} e^{-(\bar{t}-\bar{t}_0-\bar{t}')^2/2\bar{\sigma}_{et}^2} \tag{60}$$

$$f_\perp(t-t_0) = K_\perp \int_{-\infty}^{\infty} d\bar{t}' \frac{1}{(\bar{t}'^2+1)^{3/2}} \frac{1}{\sqrt{2\pi}\bar{\sigma}_{et}} e^{-(\bar{t}-\bar{t}_0-\bar{t}')^2/2\bar{\sigma}_{et}^2} \tag{61}$$

$$K_\parallel = \frac{e^2}{4\pi\varepsilon_0 (v_0 \gamma t_r)^2} \boldsymbol{r}_{i,j} \cdot \hat{\boldsymbol{e}}_z = \frac{e^2}{4\pi\varepsilon_0} \frac{\boldsymbol{r}_{i,j} \cdot \hat{\boldsymbol{e}}_z}{r_{0\perp}^2} \tag{62}$$

$$K_\perp = \frac{e^2}{4\pi\varepsilon_0 (v_0 \gamma t_r)^2} \boldsymbol{r}_{i,j} \cdot \hat{\boldsymbol{e}}_r = \frac{e^2}{4\pi\varepsilon_0} \frac{\boldsymbol{r}_{i,j} \cdot \hat{\boldsymbol{e}}_r}{r_{0\perp}^2} \tag{63}$$

Substituted in (58), this already indicates dependence of the transition probability amplitude on the QEW dimension.

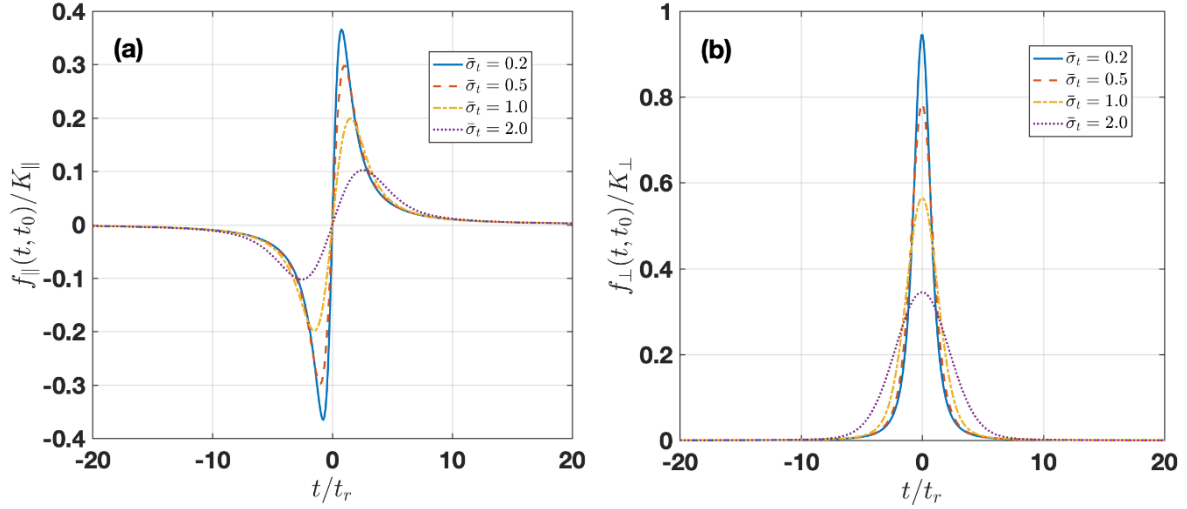

Fig. 2. The weighed interaction probability $f(t, t_0)$ for $M_{i,j\parallel}(z)$ and $M_{i,j\perp}(z)$.

Now we proceed in calculating the TLS transition amplitude (58) for the case of small change in the TLS probability amplitude during the interaction:

$$C_j(t) = C_j(t_0) \tag{64}$$

$$C_i(t_0^+) = C_i^{(0)}(t_0^-) + \Delta C_i \tag{65}$$

$$\Delta C_i = \frac{1}{i\hbar} C_j(t_0) \int_{t_0^-}^{t_0^+} e^{-i\omega_{i,j}t} f(t-t_0) dt. \tag{66}$$

Replacing the interaction over the interaction time with a Fourier transform:

$$F(\omega) = \mathcal{F}\{f(t-t_0)\} = \int_{-\infty}^{\infty} e^{i\omega t} f(t-t_0) dt, \tag{67}$$

$$\Delta C_i = \frac{1}{i\hbar} C_j(t_0) F(-\omega_{i,j}) \tag{68}$$

where $f(t-t_0)$ is the combined probability function (56). We evaluate (66) by substitution of (56) in (67) and changing the integrations order:

$$\begin{aligned}
F(-\omega_{i,j}) &= \frac{1}{v} \int_{-\infty}^{\infty} dt\, e^{-i\omega_{i,j}t} \int_{-\infty}^{\infty} dz\, M_{i,j}(z) f_{et}(t-t_0-z/v) \\
&= \frac{1}{v} \int dz\, e^{i\omega_{i,j}(t_0+z/v)} M_{i,j}(z) F_{et}(-\omega_{i,j}) \\
&= \frac{1}{v} e^{i\omega_{i,j}t_0} \tilde{M}_{i,j}\left(\frac{\omega_{i,j}}{v}\right) F_{et}(-\omega_{i,j})
\end{aligned} \tag{69}$$

For a Gaussian QEW:

$$f_{et} = \frac{1}{\left(2\pi\sigma_{et}^2\right)^{1/2}} e^{-t^2/2\sigma_{et}^2} \tag{70}$$

$$F(\omega_{i,j}) = e^{-\omega_{i,j}^2 \sigma_{et}^2/2} \tag{71}$$

$$\Delta C_i = \frac{1}{i\hbar v} C_j(t_0) e^{i\omega_{i,j} t_0} \tilde{M}_{i,j}\left(\frac{\omega_{i,j}}{v_0}\right) e^{-\omega_{i,j}^2 \sigma_{et}^2/2} \tag{72}$$

Similarly to (38):

$$P_i(t_0^+) = \left|C_i(t_0^-) + \Delta C_i\right|^2 = P_i^{(0)} + \Delta P_i^{(1)} + \Delta P_i^{(2)} \tag{73}$$

$$P_i^{(0)} = \left|C_i^{(0)}(t_0^-)\right|^2 \tag{74}$$

$$\Delta P_i^{(1)} = 2\,\mathrm{Re}\left[C_i^{(0)*}(t_0^-)\Delta C_i\right] \tag{75}$$

$$\Delta P_i^{(2)} = |\Delta C_i|^2 \tag{76}$$

For finite size QEW:

$$\Delta P_i^{(2)}(t_0^+) = \frac{1}{\hbar^2 v_0^2}\left|C_j(t_0)\tilde{M}_{i,j}\left(\frac{\omega_{i,j}}{v_0}\right)\right|^2 e^{-\Gamma^2}$$

$$P_2(t_0^+) = \Delta P_2^{(2)}(t_0^+) = \frac{1}{\hbar^2 v_0^2}\left|\tilde{M}_{i,j}(p_{rec})\right|^2 e^{-\Gamma^2} \tag{77}$$

Where the second equation corresponds to TLS excitation from ground state: $i = 2, j = 1, C_1(t_0^-) = 1$.

In the case of TLS excitation from a superposition state, the first order transition term (75) is:

$$\Delta P_i^{(1)}(t_0^+) = \frac{2}{\hbar v_0}\,\mathrm{Re}\left[C_i^{(0)*}(t_0)C_j^{(0)}(t_0)e^{-i\omega_{i,j}t_0}\tilde{M}_{i,j}\left(\frac{\omega_{i,j}}{v_0}\right)/i\right]e^{-\Gamma^2/2}$$

$$= \frac{2}{\hbar v_0}\left|\tilde{M}_{i,j}(p_{rec})\right|\left|C_i^{(0)*}(t_0)C_j^{(0)}(t_0)\right|e^{-\Gamma^2/2}\sin\varsigma \tag{78}$$

These expressions are consistent with Eqs. 42, 43, 47 in the limit of short interaction time (50), and manifest the wavepacket size dependence of the transition probabilities through the parameter $\Gamma$(48) in the near-point-particle parameters regime. The probabilistic model approximation is presumed to apply in the short QEW regime (50), and therefore (77) is not inconsistent with (42), (43) that indicate a wavepacket-independent finite contribution of the second order term. On the other hand, both (78) and (47) remarkably predict the same wavepacket-dependent decay and same relative wavepacket arrival time phase-match dependence of the incremental transition probability $\Delta P_i^{(1)}$ for a superposition state. The dependence of (78) on the resonant phase-match timing of the short interaction impulse due to the QEW arrival relative to the dipole oscillation phase, is indicative of a possible coherent interaction enhancement by multiple electrons with correlated arrival timing.

## 4. Numerical solution and verification of analytical approximations for a single QEW interaction.

In order to check the validity of the analytical approximations, we have developed two kinds of numerical computation codes for solving the FEBERI problem of interaction between a single finite size QEW and a TLS at any initial state. Extension to computation of FEBERI with modulated QEWs and with multiple correlated QEWs will be reported elsewhere. The numerical computation methods are described in Appendix E.

The computation examples of the FEBERI effect were performed for a model of a Gaussian QEW, and were studied as a function of its size $\sigma_{et}$, in order to examine the claimed dependence of the interaction on the wavepacket shape, and delineate the transition from the quantum-wave-like limit ($\sigma_{et} > T_{2,1} = 2\pi/\omega_{21}$) to the point-particle –like limit ($\sigma_{et} < T_{2,1} = 2\pi/\omega_{21}$) of the QEW. In order to focus on this parameter scaling, the computation parameters in the examples shown here, are all for a fixed tr and $\sigma_{et} > t_r$ . In all the current examples the dipole polarization was taken to be transverse (A4, A6). The parameters used in the examples are typical to electron microscope PINEM-kind experiments [8]:

Table 1: Computation parameters

| Beam Energy | $\mathcal{E}_0 = 200\text{keV}$  ($\gamma = 1.4$) |
|---|---|
| Free electron impact parameter (transit time) | $r_\perp = 2.4 nm$  ($t_r = r_\perp/c\beta\gamma = 6 atS$) |
| TLS energy gap (transition frequency) | $E_{2,1} = 2 eV$  ($\omega_{2,1} = 3\times 10^{15}\ rad/S$) |
| Dipole moment | $\mu_{i,j}^\perp = 5 Debye$ |

Figure 3 displays the time dependence of the transition probability of a TLS, starting from ground state, and the corresponding energy decrement of the free electron, demonstrating maintenance of energy conservation $E_F(t) - E_{F,in} + \hbar\omega_{12} P_2 = 0$ throughout the process. The interaction time depends on the QEW size $\sigma_{et}$, but the final result of upper level occupation probability after interaction is found to be independent of $\sigma_{et}$. This is in good match with the analytical result (43) marked of by a dot in the figure, and also consistent with (77) within its validity range $\Gamma < 1$. Simulations showed that the post interaction transition probability to the upper level from ground state stays constant also in the range $\Gamma > 1$ where the approximation (43) holds and the approximation leading to the decay prediction of (77) does not. Indeed, this result seems to be agreeable also with the philosophical point of view of Born's probability interpretation of the electron wavefunction: the probability of a point-particle arrival to the TLS location is spread over a longer time when $\sigma_{et}$ is large, but it always happens at some time during the passage of the QEW, and exhibits the same inelastic scattering.

Figure 4 displays the time dependence of the transition probability of a TLS, starting from a superposition state $(|1\rangle + i|2\rangle)/\sqrt{2}$ for different values of $\sigma_{et}$. In this case, contrary to excitation from ground level, the incremental probability for transition to the upper level depends on the wavepacket size, and decreases for longer QEW. Instructively the sum of the curves show that the

conservation of energy relation of the free electron and the bound electron is not kept during the interaction: $E_F(t) - E_{F,in} + \hbar\omega_{12} P_2 \neq 0$. This can be explained by the permission of energy uncertainty at the short interaction time regime (49). The energy difference is the temporary interaction energy. The energy conservation relation is fully kept after the interaction.

The strong wavepacket-size dependent exponential decay $\exp(-\Gamma^2/2)$ predicted by both analytical approximate expressions (43) and (78) is clearly demonstrated in Fig 5 for a general superposition starting state on the equatorial of the Bloch (Poincare) sphere $(|1\rangle + e^{i\varphi}|2\rangle)/\sqrt{2}$. The dependence on the incidence matching phase $\zeta = \varphi - \omega_{21}t_0$ and on the QEW size $\sigma_{et}$, is in full conformation with the analytical approximate expressions (43), (78). Note the big (three orders of magnitude) enhancement of the maximum incremental transition probability in the limit of short QEW. This is well explained by comparison of $\Delta P_2^{(1)}$ (Eqs. 44 or 78) and $\Delta P_2^{(2)}$ (Eqs. 42 or 77), that results in:

$$\left[\Delta P_2^{(1)}\right]_{max} = \sqrt{\Delta P_2^{(2)}}$$

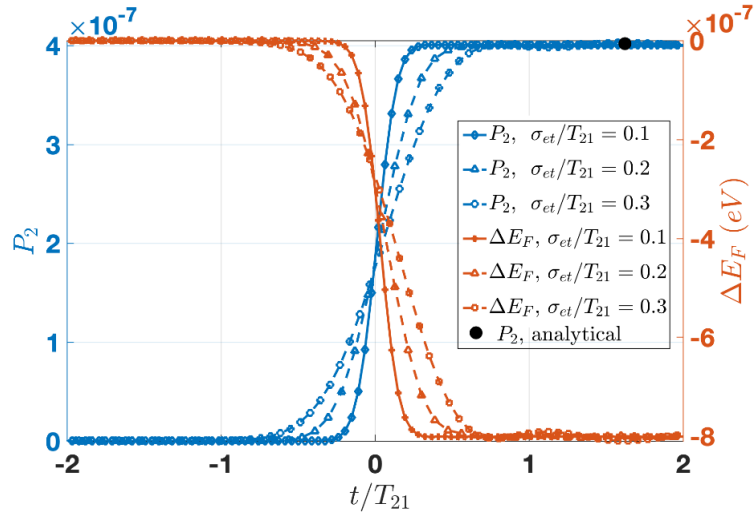

Fig. 3. Time dependence of TLS transition probability from ground state.

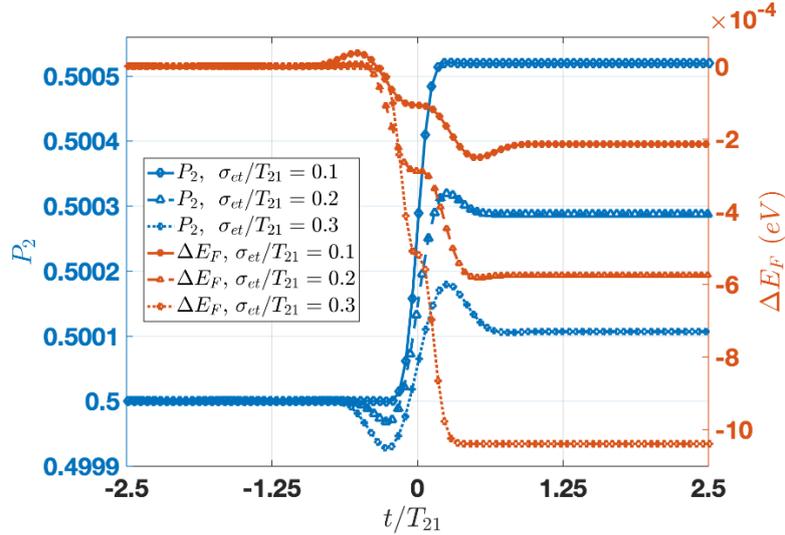

Fig. 4. Time dependence of TLS incremental transition probability starting from superposition state $(|1\rangle + e^{i\varphi}|2\rangle)/\sqrt{2}$ with the incidence relative phase $\varsigma = \omega_{21}t_0 - \varphi = \pi/2$.

In either case, the total transition increment is always: $\Delta P_2 = \Delta P_2^{(1)} + \Delta P_2^{(2)}$, namely, the incremental transition probability shown in Fig. 5 never decays to zero, but the contribution of $\Delta P_2^{(2)}$ is negligible. Figure 6 shows the same dependence in more details, displaying in the insets the sinusoidal dependence on the incidence phase and the exponential decay with the QEW size and the excellent agreement of the analytical and numerical analytical approximate expressions (43), (78). Note the big (three orders of magnitude) enhancement of the maximum incremental transition probability in the limit of short QEW.

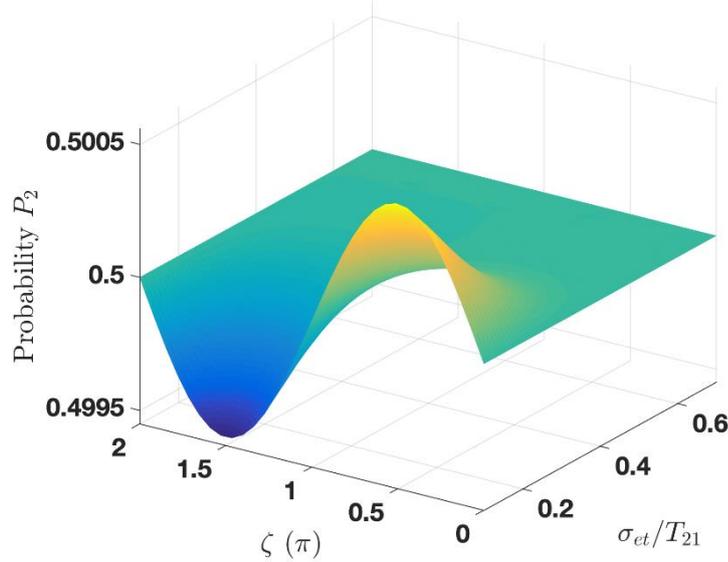

Fig. 5. Numerically computed transition probability $P_2$ as function of $\sigma_{et}$ and $\zeta$.

## 5. FEBERI interaction with a modulated QEW

Here we extend our Born's probability interpretation of the electron wavefunction to model the case of a modulated QEW. In this case we use in (57):

$$\left\langle \left|\Psi_{\kappa}^{(0)}(\mathbf{r},t)\right|^2 \right\rangle = \delta(r_\perp) f_{et}(t - t_{0K} - z/v_0) f_{mod}(t - z/v - t_L) \quad (79)$$

where the modulation function is periodic [8,18] (see Fig. 7):

$$f_{mod}(t) = f_{mod}(t + 2\pi/\omega_b) \quad (80)$$

and therefore:

$$f_{mod}(t) = \sum_{m=-\infty}^{\infty} f_m e^{im\omega_b t} \quad (81)$$

The coefficients $f_m$ were derived in [18] for the case of the wavefunction of a modulated Gaussian QEW (15), $\omega_b t_L$ is the modulation phase, determines by the modulating laser beam.

The incremental excitation probabilities (73) are derived in Append. C Explicitly:

$$\Delta P_i^{(1)} = 2\text{Re}\left\{ \frac{1}{2\pi i \hbar v_0} \tilde{M}_{i,j}\left(\frac{\omega_{i,j}}{v_0}\right) C_i^*(t_{0K}^-) C_j(t_{0K}) f_n e^{-in\omega_b t_L} e^{-(\omega_{i,j} - n\omega_b)^2 \sigma_{et}^2/2} \right\} \quad (82)$$

$$\Delta P_i^{(2)} = \left| \frac{1}{\hbar v_0} \tilde{M}_{i,j}\left(\frac{\omega_{i,j}}{v_0}\right) C_j(t_{0K}) \right|^2 |f_n|^2 e^{-(\omega_{i,j}-n\omega_b)^2 \sigma_{et}^2} \tag{83}$$

where n is the bunching frequency harmonic that matches the TLS quantum energy levels (1):

$$\omega_{i,j} = n\omega_b \tag{84}$$

Remarkably, both incremental probabilities display resonant excitation characteristics around condition (84), which would manifest the QEW modulation characteristics in a properly set experiment. Note that in a modulated QEW $\omega_{ij}\sigma_{et} > 1$, and according to (43), (47), in this range the second order incremental modulation of an unmodulated QEW is not wavepacket-dependent, and the first order increment decays. In contrast, Eqs. 82, 83 suggest resonant behavior at harmonics of the modulation frequency for the same QEWs, if modulated. This indicates a possibility for measuring the modulation features of the QEW. However, note that for single modulated QEWs, there is no enhancement of the transition probability even at resonance.

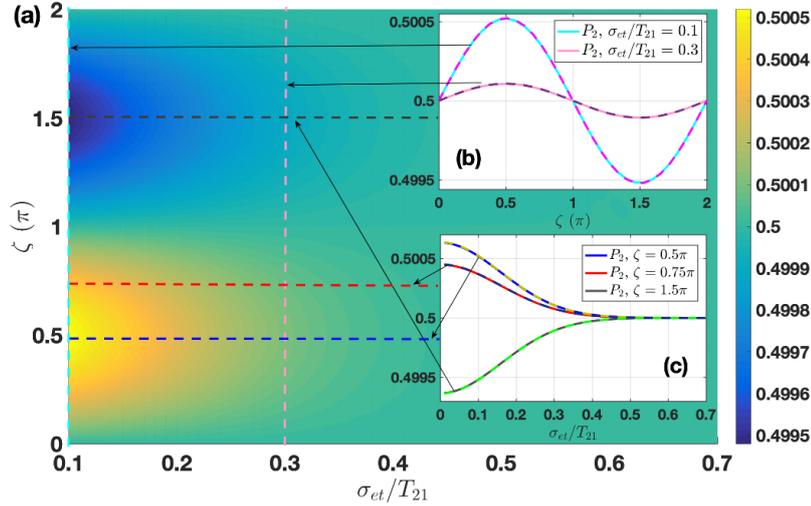

Fig. 6. (a) Numerically computed transition probability $P_2$ as function of $\sigma_{et}$ and $\varsigma$. Inset (b): Numerical (solid-lines) and analytical (dash lines) transition probability $P_2$ as function of $\zeta$, for $\sigma_{et}/T_{21} = 0.1$ and 0.3. Inset (c) Numerical (solid-lines) and analytical (dash lines) transition probability $P_2$ as function of $\sigma_{et}$, for different relative incidence phases.

### 6. FEBERI interaction with periodically bunched Multiple near-point-particle QEWs:

The FEBERI interaction with multiple QEWs is theoretically an intricate problem that involves entanglement of the free electron wavefunction with the TLS quantum states, and consequently – entanglement with all subsequent interacting electrons. We carry out such a numerical computation analysis elsewhere in a rigorous multi-particle density matrix formulation. At present, we resort again to a simple analytical approximate model, in which we extend (54), (57), (58) to multiple particles by the substitution:

$$\left\langle \left| \Psi_F^{(0)}(\mathbf{r},t) \right|^2 \right\rangle \to \sum_{K=1}^{N} \left\langle \left| \Psi_K^{(0)}(\mathbf{r},t) \right|^2 \right\rangle \tag{85}$$

$$\left\langle \left| \Psi_K^{(0)}(\mathbf{r},t) \right|^2 \right\rangle = \delta(\mathbf{r}_\perp) f_{et}(t - t_{0K} - z/v_0) \tag{86}$$

We then solve for the cumulative incremental FEBERI transition probability for the case of periodically injected near-point-particle QEWs. Under the assumption that the relaxation time of the TLS [44] is much longer than the duration of the N QEWs pulse, this results in (Appendix D):

$$P_2 = N^2 \left\{ \frac{1}{\hbar v_0} \left| \tilde{M}_{2,1}\left(\frac{\omega_{2,1}}{v_0}\right) \right| \right\}^2 e^{-\omega_{i,j}^2 \sigma_{et}^2} \tag{87}$$

This approximate result may not be rigorous in the initial stage of the multiple electrons transition buildup, if it starts from ground state. Only when N is large enough, the phase of the TLS gets established by the first near-point-particle QEWs of the train, and the subsequent QEWs then continue to build up the transitions in phase.

This case of a periodically spaced train of near-point-particle QEWs (87) may be realistic only at low (microwave or THz) frequency TLS transitions, where classical Klystron-kind electron current modulation is available. It has thus been termed a "Quantum Klystron", and analyzed in [35]. It can be comprehended as the quadratic approximation of the $\sin^2(\Omega_R t/2)$ scaling of a Rabi oscillation process, and it is the analogue of the classical bunched-particles beam superradiance effect [27]. Note that in the classical point particle limit and low (microwave) frequencies [35] high current density of the electron beam is allowed (with the limitations of beam quality and space charge effect) and there may be then multiple electrons per period, and N should be taken then as the number of electrons in the entire modulated electron beam pulse: $N = I_{mod} t_{pulse}/e$. We also point out that the case of a multiple periodic train of QEWs, is closely related to the earlier studied effect of "pulsed beam scattering" [51, 52].

It is instructive to compare the quadratic dependence of (87) on the number of QEWs -$N^2$ to the same dependence in the case of superradiance [19, 27]. In this comparison, the exponential decay factor $e^{-\omega_{ij}^2 \sigma_{et}^2}$ that originates from the finite size of the Gaussian QEW (72), is the quantum limit of the "bunching coefficient" in a bunched point-particle beam superradiance. In the case of a long wavepacket, the non-decaying probability expression for $P_2$ becomes relevant, and it would result in a multiple electrons TLS transition rate proportional to N, in analogy to the "shot-noise" spontaneous radiation emission limit of bunched-beam superradiance in the large bunching coefficient limit [19, 27].

### 7. FEBERI interaction with multiple modulation-correlated QEWs:

We here extend again our Born's probability interpretation analytical model to the case of multiple modulation-correlated QEWs. Combining the cases of the last two sections, we consider the case of multiple QEWs, all modulated at the PINEM interaction point at the level of their quantum wavefunctions [8] by the same coherent laser beam of frequency $\omega_b$ and phase $\omega_b t_L$. We use the multi-particles probability function (85), with

$$\left| \Psi_K^{(0)}(\mathbf{r},t) \right|^2 = \delta(\mathbf{r}_\perp) f_{et}(t - t_{0K} - z/v) f_{mod}(t - z/v_0 - t_L) \tag{88}$$

where $t_{0K}$ are the centroid arrival times of the envelopes of the modulated QEWs, and the modulation function, common to all QEWs is periodic in time as in (80):

$$f_{mod}(t) = f_{mod}(t + 2\pi/\omega_b) \tag{89}$$

With (81), the multi-electron probability distribution function is:

$$f(t-t_0) = \sum_{K=1}^{N} \frac{1}{v} \int dz M_{i,j}(z) f_{et}(t - t_{0K} - z/v_0) \sum_{m=-\infty}^{\infty} f_m e^{i\omega_b(t-z/v_0-t_L)} \tag{90}$$

Then, as in Append. C, substitution in (58), changing the integration order of z and t, results in:

$$C_i(t_{0N}^+) = C_i(t_0^-) + \frac{1}{i\hbar v_0} \int dz M_{i,j}(z) \sum_{K=1}^{N} C_j(t_{0K}) \sum_m \int dt\, e^{-i\omega_{i,j}t} f_{et}(t-t_{0K}-z/v_0) f_m e^{im\omega_b(t-z/v_0-t_L)}$$

$$= C_i(t_0^-) + \frac{1}{i\hbar v_0} \int dz M_{i,j}(z) e^{i\frac{\omega_{i,j}}{v_0}z} \sum_{K=1}^{N} C_j(t_{0K}) \sum_m f_m \int dt'\, e^{-i(\omega_{i,j}-m\omega_b)t'} f_{et}(t'-t_{0K}) e^{-im\omega_b t_L} \tag{91}$$

$$C_i(t_{0N}^+) = C_i(t_0^-) + \frac{1}{i\hbar v_0} \tilde{M}_{i,j}\left(\frac{\omega_{i,j}}{v_0}\right) \sum_m f_m \sum_{K=1}^{N} C_j(t_{0K}) e^{i(\omega_{i,j}-m\omega_b)t_{0K}} F_{et}(\omega_{i,j}-m\omega_b) e^{-im\omega_b t_L} \tag{92}$$

where $F_{et}(\omega_{ij} - m\omega_b) = \mathfrak{F}\{f_{et}(t)\}|_{\omega=\omega_{12}-m\omega_b}$ is the Fourier transform of the single QEW probability function. For a wide Gaussian distribution, the envelope function (70) is wide - $\sigma_{et} > 2\pi/\omega_b$, and therefore, the spectral function

$$F(\omega_{i,j} - m\omega_b) = e^{-(\omega_{i,j}-m\omega_b)^2 \sigma_{et}^2/2} \tag{93}$$

is a narrow function around a harmonic m=n that is resonant at the transition frequency:

$$\omega_{i,j} = n\omega_b \tag{94}$$

Take i=2, j=1 (upper and lower levels), then with the approximation of small change in the amplitude:

$$C_1(t_{0K}) \cong C_1(t_0^-) = \text{const} \cong 1, \quad |C_2(t_0^-)| \ll 1 \tag{95}$$

$$C_2(t_{0N}^+) \cong \frac{1}{i\hbar v_0} \tilde{M}_{i,j}\left(\frac{\omega_{i,j}}{v_0}\right) \sum_m f_m \sum_{K=1}^{N} e^{i(\omega_{i,j}-m\omega_b)t_{0K}} e^{-im\omega_b t_L} e^{-(\omega_{i,j}-m\omega_b)^2 \sigma_{et}^2/2} \tag{96}$$

This averages to zero for random arrival times $t_{0K}$ of the wavepacket centroids, except at the resonance condition (94), where independently of the arrival times $t_{0K}$:

$$\left. C_2(t_{0N}^+) \right|_{\omega_{ij}=m\omega_L} \approx N \frac{1}{i\hbar v_0} \tilde{M}_{2,1}\left(\frac{\omega_{2,1}}{v_0}\right) f_n e^{-in\omega_b t_L} \tag{97}$$

$$P_2 = N^2 \left\{ \frac{1}{\hbar v_0} \left| \tilde{M}_{2,1}\left(\frac{\omega_{2,1}}{v_0}\right) f_n \right| \right\}^2 \tag{98}$$

This expression, explicitly manifests the $N^2$ scaling buildup of the upper quantum level probability in the case of multiple modulation-correlated QEWs, similarly to the case of periodically modulated point particles (section 6) and in analogy to superradiance of bunched particles [27]. Also note that as in the derivation of Eq. 87, this approximate derivation result may not be rigorous in the initial stage of the multiple electrons transition buildup and applies for $N \gg 1$, when the phase of the TLS is established. Only when N is large enough the phase of the TLS gets established.

## 8. Simulation of FEBERI with multiple correlated QEWs

To affirm the validity of the FEBERI effect and its quadratic scaling $N^2$ with the number of modulation-correlated QEWs, we simulate the TLS transitions and the quantum levels population dynamics in a model where the electron interaction times are determined by Born's combined probability function (85,88,89), that accounts for both the classical random injection times of the centroids of the QEWs $t_{0K}$ and the quantum probability timings of the electrons arrival corresponding to the probability function $f_{mod}(t - z/v_0 - t_L)$ common to all modulation-correlated QEWs. It has the same modulation phase for all electrons - $\omega_b t_L$ (the phase of the modulating laser).

We use for our simulations the parameters of [8] corresponding to a PINEM experiment with a 200keV electron beam. The modulation probability function is the squared absolute value of the QEW wavefunction after PINEM laser interaction (15), evaluated at an optimal post-interaction drift time $t_D = T_b/2(\Delta p_{mod}/p_0)$, where $\Delta p_{mod}$ is the momentum modulation amplitude at the PINEM interaction point [18]. This distribution is shown in Fig. 4 for the parameters of [8, 18]. The density modulation bunches in this case are of attosecond time duration, much shorter than the optical frequency period of the modulation.

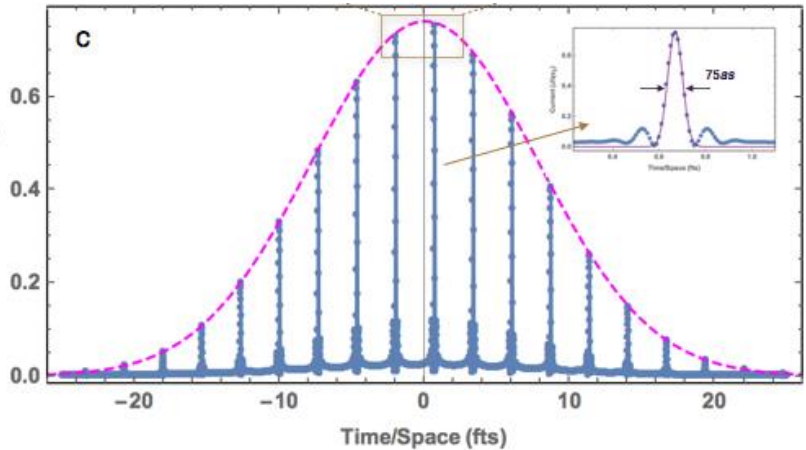

Fig. 7. The density modulated QEW distribution displayed in terms of time $(t - t_{0K})$ relative to $t_{0K}$- the centroid arrival time of electron K (from [8,18]).

Therefore, we assert that even in case that the wavepacket duration $\sigma_t$ of an unmodulated QEW is long relative to the TLS transition resonant frequency $\omega_{2,1}$, not satisfying the near-point-particle regime condition (50), the micro-bunches of such a tightly bunched modulated QEWs may still operate in this regime.

Contrary to the analytical approach in section 7, in the simulations we calculate the accumulative dynamics of the TLS transitions due to multi-electron interactions by calculating separately the transition dynamics of the single near-point-particle QEWs (56, 57) and adding them in sequence. Guided by the probabilistic interpretation of (53), we use the probability distribution function (88) only as the algorithm for determining the interaction time $t_K$ of a modulated QEW of centroid arrival time $t_{0K}$ (see Fig. 7). For simplicity we make the approximation of very tight density bunching (see [8,9,12], where attoSecond short bunches were demonstrated, while the optical period $T_b = 2\pi/\omega_b$ was of the order of femtoSeconds). Therefore, under the assumption that the QEW's envelope duration $\sigma_{et}$ is longer than the optical period, the arrival times of the electrons are determined by the modulation function peaks of the quantum probability distribution displayed in Fig. 7:

$$t_K = t_{0L} + n_K T_b, \qquad (99)$$

where $n_K$ is an ascending series of randomly spaced integers K=1..N. The sample electrons K that contribute to the transition amplitude dynamics through Eq. 57 are only the ones that their arrival times $t_K$ fall within the range of the QEW envelope width (see Fig. 7) of any QEW that reaches at a centroid random time $t_{0K}$. This is almost the only role of $t_{0K}$, and the interaction time (99) is determined primarily by the micro-bunch peaks (Fig. 7) of the Born quantum probability distribution function (88). In this picture, Eq. 99 represents a train of near point-particle wavepackets spaced with integral multiple optical periods, such that only one electron is counted within the envelope of each modulated QEW. Therefore, as predicted in the analytical calculation of Section 7, it would be expected that simulation of such a train of phase-correlated particles would resonantly buildup the transition to the upper level when the modulation frequency $\omega_b$ is a sub-harmonic of the transition frequency:

$$\omega_{2,1} = n\omega_b, \qquad (100)$$

The transition probability of a single electron wavepacket was calculated by solving the coupled equations (57) (i, j = 1, 2) for a QEW in the near-point-particle limit for the parameters given in Table 1. This is shown in Fig. 8.

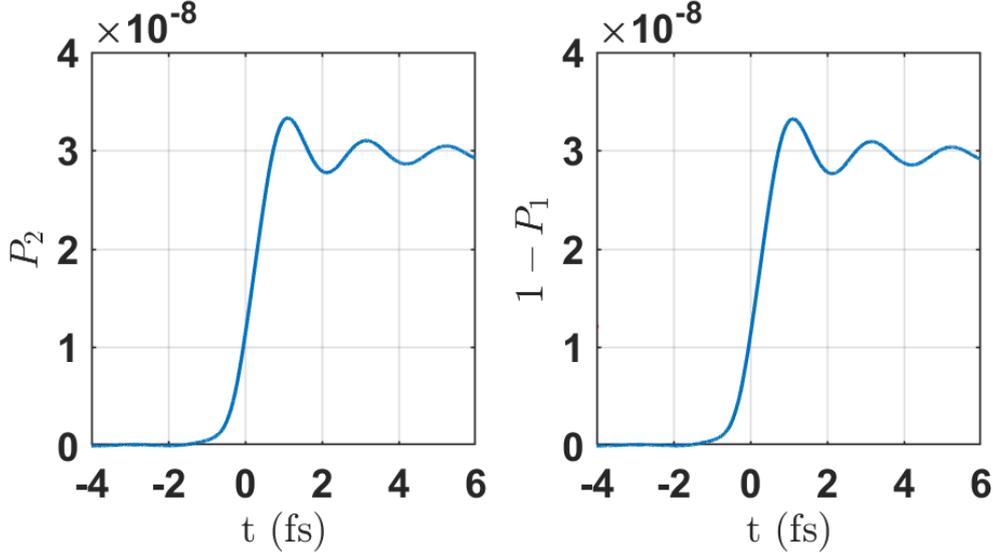

Fig. 8. Numerical solution of the coupled equations (57) for the transition between the TLS quantum states. $P_1$ and $P_2$ represent the occupation probability of the ground state and upper state respectively, satisfying the relation $P_1 + P_2 = 1$. The parameters of the exciting QEW are in the near-point-particle regime.

Figure 9 shows simulation of the buildup of the TLS upper level probability, solving (57) with $N_1 = 20$ particles arriving at times (99), where $n_K$ is a random number (blue curve). The growth is evidently quadratic, $P_2 \propto N^2$, as claimed. For comparison, we show in the upper (magenta) curve the case where instead of (99), $t_K$ is taken to be entirely random. The growth rate is linear, and the upper TLS level arrives at the same excitation level only with $N_2 = N_1^2 = 400$.

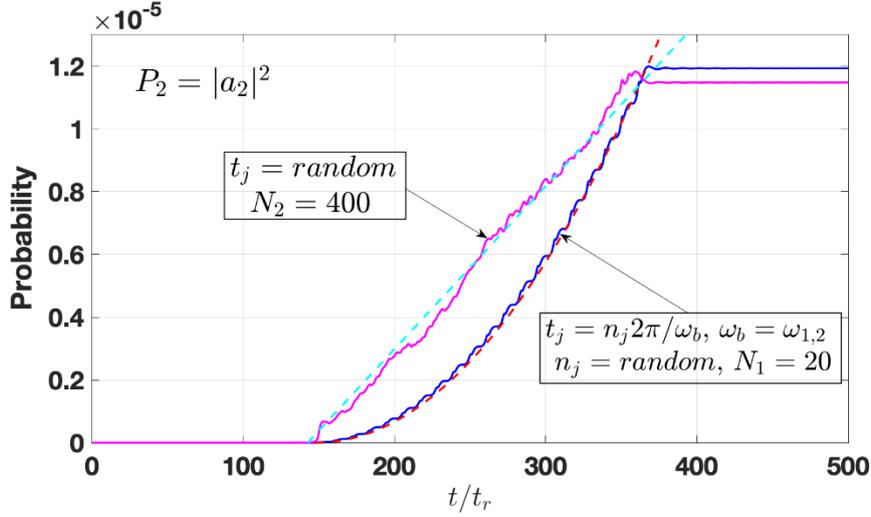

Fig. 9. Simulation results of the upper level probability buildup by electrons arriving to the interaction point at random (magenta) and by electrons arriving in-phase with the modulating laser modulo the bunching period $T_b = 2\pi/\omega_b$ (99) at the resonance condition (100) (blue). The cyan-dash and red-dash curves are the linear and quadratic curve-fittings, respectively.

**Conclusion**

We studied the interaction between free electrons and a bound electron (FEBERI) in a framework of a comprehensive quantum model, in which the bound electron is represented by a two-level system (TLS) and the free electrons are represented as quantum electron wavepackets (QEW). The electric dipole interaction of the two-body system was solved in terms of the Schrodinger equation, extended to the relativistic regime. Since we start from a wavepacket model for the free electron, the analysis can provide two limits of interaction: wavelike and point-particle-like interactions, manifesting the wave-particle duality nature of quantum mechanics and the transition from quantum to classical point-particle presentations. This observation is similar to the analogous cases of QEWs interactions with light: stimulated radiative interaction and superradiance [16-19]. The results show that the electron wavepacket dimension and its density modulation characteristics [8] are physically observable parameters that can be measured in appropriately designated experimental setups of electron interaction with matter or light. However, since the QEW dimensions and modulation characteristics are defined only in terms of expectation values, this assertion applies only to measurement of interaction with multiple identical QEWs.

Contrary to the previous semiclassical model used in [1], the analysis in this paper is fully quantum mechanical, rigorously applied to the case of single electrons, modeled as quantum wavepackets. Two different analytical approximation analyses were derived for TLS excitations from ground state and from a general superposition state, and they were shown to agree well, in their regimes of validity, with the exact numerical computation solutions of Schrodinger equation. In the case of a superposition TLS state, we find wavepacket size-dependent significant enhancement of TLS quantum transitions in the short (near-point-particle) limit of the QEW that would come into expression in laboratory measurement of EELS and Cathodoluminescence. Furthermore, the transition probability to an upper or lower quantum level of the TLS depends on the arrival time of the near-point-particle QEW (typically sub-ftSec) relative to the phase of the TLS in its Bloch sphere representation. This observation may lead to potential applications in new electron microscopy atomic-scale probing of quantum excitations in matter, and particularly may be usable for an important application of coherent control of Q-bits and quantum emitters.

In the framework of a rigorous quantum mechanical relativistic model of single QEW interaction with a TLS, we show that the results of our two analytical approximation models of quantum wavepacket-size-dependent electron-matter interaction, are in good match with the results of exact numerical computation solution of Schrodinger equation. These results are in good agreement with the Born interpretation of the electron quantum wavefunction (in the context of ensemble average). Based on this observation, we adopted a Born probability inspired hypothesis that the demonstrated quantum wavepacket shape dependence would apply also to a case of optical frequency modulated QEWs. We have shown (section 5) that such a model results in prediction of a laboratory measureable resonant FEBERI effect at the condition that the TLS transition frequency is a harmonic of the modulation frequency of the QEWs: $n\omega_b = \omega_{2,1}$. This resonant excitation may be observable in measurement of cathodoluminescence of the TLS and EELS of the free electrons.

A further extension of the Born interpretation hypothesis model of FEBERI interaction to the case of multiple electrons (in section 6), reveals a possibility for enhanced resonant transition probability buildup of the upper TLS quantum level in proportion to the number of electrons squared - N2. This would happen when the QEWs are short, namely in the near-point-particle limit of the QEW, and their wavefunction centroids are periodically density-bunched at a subharmonic of the TLS quantum transition frequency under the condition that the duration of the electron beam pulse is shorter than the relaxation time of the TLS. This case corresponds to a "quantum klystron" [35, 51, 52], where the QEWs interaction can be described in terms of periodic bunching of classical single particles, analogous to classical superradiance of a bunched particle beam [26, 27].

Finally, the Born hypothesis model was further extended (in sections 7, 8) to the case of a pulse of randomly injected QEWs that are, however, modulated by the same laser beam, so that the expectation value of the density distribution of each QEW is modulated at the laser optical frequency (Fig. 7) and their modulation phases are correlated. Under the hypothesis that the Born interpretation model applies to multiple modulation-correlated QEWs, we show that also in this case, similar quadratic N2 buildup of the upper TLS level is expected when the QEWs are tightly modulated, and a harmonic of their common modulation frequency is resonant with the TLS quantum transition frequency. This coherent transitions buildup can be explained as a result of arrival times of a train of quantum-probability-determined "point-particles" in-phase with the TLS transition frequency, even though the centroids of the QEWs arrive at random. Such a coherent buildup process of the quantum transitions of the TLS would be expressed and measureable in the lab in an ultrafast Transmission Electron Microscope setting as an enhanced cathodoluminescence effect, and a corresponding EELS spectrum change of the electron beam.

The analysis here, though comprehensively quantum-mechanical for the case of a single QEW, is still incomplete for the case of multiple particle FEBERI. This process requires further elaboration, considering the entanglement that is established in the process between the QEW and the TLS quantum state, and further with the subsequent incoming electrons. This kind of analysis is under way elsewhere in terms of density matrix formulation. These predictions of the FEBERI process call for experimental verification, requiring advancement of the technological state of the art of electron microscopy. Shaping the wavepacket length in the range of an optical period requires development of filtering and wavepacket compression techniques [48, 49, 55]. One must be also concerned about mitigation of the electron beam deterioration due to Coulomb interaction between the electrons [38] (e.g. by use of a high repetition rate mode locked laser [39]). One must be aware also that the expected probabilities of transition per single TLS in the given examples are very small, and experimental verification would probably require use of multiple TLS schemes, such as superradiant cathodoluminescence effect [50]. Much of the technical difficulties would be mitigated in scenarios of low frequency (microwave, THz) TLS transition frequencies, as in [55].

Finally, we stress that the present single QEW analysis of the FEBERI interaction, and the conclusions about the measurability of the QEW dimensions, are based on the expectation value of the density probability distribution, and there is an explicit assumption there that the measurements are done with an ensemble of properly prepared identical multiple electrons. We mention however, the prevalent research interest in non-projective direct measurability of single particle wavefunctions using weak measurement [40, 41] or protective measurement [42, 43] schemes. This aspect, representing an alternative realistic interpretation of the QEW, is not covered in the present article.

**Appendix A: Coulomb Interaction Matrix element**

Here we evaluate the momentum space representations of the interaction matrices of the longitudinally and transversely aligned dipoles (25).

For the longitudinally aligned dipole:

$$M_{i,j,\parallel}(z) = \frac{e^2 r_{i,j}}{4\pi\varepsilon_0} \frac{\gamma z}{(\gamma^2 z^2 + r_\perp^2)^{3/2}} \tag{A1}$$

Then:

$$\widetilde{M}_{i,j,\parallel}(p) = \int dz\, M_{i,j,\parallel}(z) e^{-ipz/\hbar} = \frac{e^2 r_{i,j}}{4\pi\varepsilon_0 \gamma^2} \int dz\, \frac{z}{(z^2 + r_\perp^2/\gamma^2)^{3/2}} e^{-ipz/\hbar}$$

$$= \frac{e^2 r_{i,j}}{4\pi\varepsilon_0 \gamma^2} \int dz\, \frac{z}{(z^2 + r_\perp^2/\gamma^2)^{3/2}} \left[-i \sin\left(\frac{p}{\hbar} z\right)\right] \tag{A2}$$

Using a relation from an integrals table [47]:

$$\widetilde{M}_{i,j,\parallel}(p) = -i \frac{e^2 r_{i,j}}{4\pi\varepsilon_0 \gamma^2} \frac{p}{\hbar} K_0\left(\frac{p r_\perp}{\hbar \gamma}\right) \tag{A3}$$

For the transversely aligned dipole:

$$M_{i,j,\perp}(z) = \frac{e^2 r_{i,j}}{4\pi\varepsilon_0} \frac{\gamma r_\perp}{(\gamma^2 z^2 + r_\perp^2)^{3/2}} \tag{A4}$$

So the Fourier transformation of the function is

$$\widetilde{M}_{i,j,\perp}(p) = \int dz\, M_{i,j,\perp}(z) e^{-ipz/\hbar} = \frac{e^2 r_{i,j} r_\perp}{4\pi\varepsilon_0 \gamma^2} \int dz\, \frac{1}{(z^2 + r_\perp^2/\gamma^2)^{3/2}} e^{-ipz/\hbar} =$$

$$= \frac{e^2 r_{i,j} r_\perp}{4\pi\varepsilon_0 \gamma^2} \int dz\, \frac{1}{(z^2 + r_\perp^2/\gamma^2)^{3/2}} \cos\left(\frac{p}{\hbar} z\right) \tag{A5}$$

Using [47]:

$$\widetilde{M}_{i,j,\perp}(p) = \frac{e^2 r_{i,j}}{4\pi\varepsilon_0 \gamma^2} \frac{1}{\sqrt{2\pi}} \frac{p}{\hbar} K_1\left(\frac{p r_\perp}{\hbar \gamma}\right) \tag{A6}$$

# Appendix B: Calculate $I = \int c_p^{(0)*} c_{p-p_{rec}}^{(0)} dp$

We calculate the coefficient $I(p)$ of Eq. 45 for a finite Gaussian QEW using Eq. 12, and assuming for simplicity that the QEW arrives at its longitudinal waist (no chirp) - $t_D = L_D/v_0 = 0$, $\tilde{\sigma}_p = \sigma_{p0}$, arriving to the interaction point at tome $t_0$: $c_p^{(0)}(t_0) = \frac{1}{(2\pi\sigma_{p0}^2)^{1/4}} e^{-(p-p_0)^2/4\tilde{\sigma}_p^2} e^{-iE_p t_0/\hbar}$

$$I(p_{rec}) = \frac{1}{(2\pi\sigma_{p0}^2)^{1/2}} \int_p dp\, e^{-\frac{(p-p_0)^2}{4\sigma_{p0}^2}} e^{-\frac{(p-p_0-p_{rec})^2}{4\sigma_{p0}^2}} \tag{B1}$$

$$I(p_{rec}) = \frac{1}{(2\pi\sigma_{p0}^2)^{1/2}} \int_p dp\, e^{-\frac{(p-p_0)^2}{4\sigma_{p0}^2}} e^{-\frac{(p-p_0-p_{rec})^2}{4\sigma_{p0}^2}} e^{-i(E_p - E_{p-p_{rec}})t/\hbar} \tag{B2}$$

We substitute $E_p - E_{p-p_{rec}} = p_{rec}\, \partial E_p/\partial p = p_{rec} v_0 = \hbar\omega_{i,j}$, and complete the polynomial to a square:

$$\frac{1}{(2\pi\sigma_{p0}^2)^{1/2}} \left[\int_p dp\, e^{-(p-p_0-p_{rec}/2)^2/2\sigma_{p0}^2}\right] e^{-\frac{1}{2}(p_{rec}/2\sigma_{p0})^2} e^{-i\omega_{i,j}t_0} = 1 \cdot e^{-\frac{1}{2}(p_{rec}/2\sigma_{p0})^2} e^{-i\omega_{i,j}t_0} \tag{B3}$$

Note:

With $p_{rec} = \hbar\omega_{1,2}/v_0$, $\sigma_{z0} = \hbar/2\sigma_{p0}$, $\sigma_{t0} = \sigma_{z0}/v_0$, we define (3), in analogy to the case of QEW interaction with light [16]:

$$\Gamma = \frac{p_{rec}}{2\sigma_{p0}} = \frac{\hbar\omega_{1,2}}{2v_0\sigma_{p0}} = \omega_{1,2}\sigma_{t0} = 2\pi\frac{\sigma_{z0}}{\beta\lambda_{1,2}} \tag{B4}$$

then

$$I(p_{rec}) = e^{-\Gamma^2/2} e^{-i\omega_{i,j}t_0} \tag{B5}$$

## Appendix C: FEBERI Interaction with a modulated QEW

Starting from (79), (81)

$$\left\langle \left| \Psi_K^{(0)}(\mathbf{r},t) \right|^2 \right\rangle = \delta(\mathbf{r}_\perp) f_{et}(t - t_{0K} - z/v) f_{mod}(t - z/v_0 - t_L) \tag{C1}$$

$$f_{mod}(t) = \sum_{m=-\infty}^{\infty} f_m e^{im\omega_b t} \tag{C2}$$

The distribution function (56) becomes

$$f(t - t_0) = \frac{1}{v_0} \int dz M_{i,j}(z) f_{et}(t - t_{0K} - z/v_0) \sum_{m=-\infty}^{\infty} f_m e^{i\omega_b(t - z/v_0 - t_L)} \tag{C3}$$

We substitute this probability distribution of a modulated QEW in (58) and change order of integrations:

$$C_i(t_0^+) = C_i(t_0^-) + \frac{1}{2\pi i \hbar v_0} \int dz M_{i,j}(z) C_j(t_{0K}) \sum_m \int dt\, e^{-i\omega_{i,j}t} f_{et}(t - t_{0K} - z/v_0) f_m e^{im\omega_b(t - z/v_0 - t_L)} \tag{C4}$$

With change of variables $t' = t - z/v_0$,

$$\begin{aligned} C_i(t_0^+) &= C_i(t_0^-) + \frac{1}{2\pi i \hbar v_0} \int dz M_{i,j}(z) e^{i\frac{\omega_{i,j}}{v_0}z} C_j(t_{0K}) \sum_m f_m \int dt'\, e^{-i(\omega_{i,j} - m\omega_b)t'} f_{et}(t' - t_{0K}) e^{-im\omega_b t_L} \\ &= C_i(t_{0K}^-) + \frac{1}{2\pi i \hbar v_0} \tilde{M}_{i,j}\!\left(\frac{\omega_{i,j}}{v_0}\right) C_j(t_{0K}) \sum_m f_m e^{i(\omega_{i,j} - m\omega_b)t_{0K}} F_{et}(\omega_{i,j} - m\omega_b) e^{-im\omega_b t_L} \end{aligned} \tag{C5}$$

We calculate the incremental transition probabilities (75), (76):

$$\begin{aligned} \Delta P_i^{(1)} &= 2\operatorname{Re}\!\left[ C_i^{(0)*}(t_0^-) \Delta C_i \right] = \\ &\quad 2\operatorname{Re}\!\left\{ \frac{1}{2\pi i \hbar v_0} \tilde{M}_{i,j}\!\left(\frac{\omega_{i,j}}{v}\right) C_i^*(t_{0K}^-) C_j(t_{0K}) \sum_m f_m e^{i(\omega_{i,j} - m\omega_b)t_{0K}} F_{et}(\omega_{i,j} - m\omega_b) e^{-im\omega_b t_L} \right\} \end{aligned} \tag{C6}$$

If the Gaussian distribution (70) is a wide function $\sigma_{et} > 2\pi/\omega_b$, then the spectral function

$$F(\omega_{i,j} - m\omega_b) = e^{-(\omega_{i,j} - m\omega_b)^2 \sigma_{et}^2/2} \tag{C7}$$

is a narrow function around a harmonic m=n that is resonant with the transition frequency:

$$\omega_{i,j} = n\omega_b \tag{C8}$$

Then only one harmonic - n can excite resonantly the transition in (C6):

$$\Delta P_i^{(1)} = 2\operatorname{Re}\!\left\{ \frac{1}{2\pi i \hbar v_0} \tilde{M}_{i,j}\!\left(\frac{\omega_{i,j}}{v_0}\right) C_i^*(t_{0K}^-) C_j(t_{0K}) f_n e^{i(\omega_{i,j} - n\omega_b)t_{0K}} e^{-in\omega_b t_L} e^{-(\omega_{i,j} - n\omega_b)^2 \sigma_{et}^2/2} \right\} \tag{C9}$$

under the condition that it is phase-matched to the phase of the initial dipole moment $C_i^*(t_{0K}^-) C_j(t_{0K})$.

Likewise:

$$\Delta P_i^{(2)} = \left|\Delta C_i\right|^2 = \left|\frac{1}{2\pi\hbar}\tilde{M}_{i,j}\left(\frac{\omega_{i,j}}{v_0}\right)C_j(t_{0K})\right|^2 \left|\sum_m f_m e^{i(\omega_{i,j}-m\omega_b)t_{0K}} F_{et}(\omega_{i,j}-m\omega_b)e^{-im\omega_b t_L}\right|^2 =$$

$$= \left|\frac{1}{2\pi\hbar v_0}\tilde{M}_{i,j}\left(\frac{\omega_{i,j}}{v}\right)C_j(t_{0K})\right|^2 |f_m|^2 \left|F_{et}(\omega_{i,j}-n\omega_b)\right|$$

(C10)

$$\Delta P_i^{(2)} = \left|\frac{1}{2\pi\hbar v_0}\tilde{M}_{i,j}\left(\frac{\omega_{i,j}}{v_0}\right)C_j(t_{0K})\right|^2 |f_n|^2 e^{-(\omega_{i,j}-n\omega_b)^2 \sigma_{et}^2}$$

(C11)

Both incremental probabilities are dependent on the QEW shape and modulation features.

## Appendix D: FEBERI interaction with multiple QEWs

Starting from equations (85), (86):

$$\left\langle \left|\Psi_F^{(0)}(\mathbf{r},t)\right|^2 \right\rangle \to \sum_{K=1}^{N} \left\langle \left|\Psi_K^{(0)}(\mathbf{r},t)\right|^2 \right\rangle \tag{D1}$$

$$\left\langle \left|\Psi_K^{(0)}(\mathbf{r},t)\right|^2 \right\rangle = \delta(\mathbf{r}_\perp) f_{et}(t - t_{0K} - z/v_0) \tag{D2}$$

we substitute the N particles probability function

$$f(t-t_0) = \sum_{K=1}^{N} \frac{1}{v} \int dz\, M_{i,j}(z) f_{et}(t - t_{0K} - z/v_0) \tag{D3}$$

in (58), and changing order of integrations in z and t:

$$C_i\left(t_{0N}^+\right) = C_i\left(t_0^-\right) + \frac{1}{2\pi i \hbar v_0} \int dz\, M_{i,j}(z) \sum_{K=1}^{N} C_j(t_{0K}) \int dt\, e^{-i\omega_{i,j}t} f_{et}(t - t_{0K} - z/v_0) \tag{D4}$$

With change of variables $t' = t - z/v_0$,

$$\begin{aligned}
C_i\left(t_{0N}^+\right) &= C_i\left(t_0^-\right) + \frac{1}{2\pi i \hbar v_0} \int dz\, M_{i,j}(z) e^{i\frac{\omega_{i,j}}{v}z} C_j(t_{0K}) \sum_{K=1}^{N} \int dt'\, e^{-i\omega_{i,j}t'} f_{et}(t' - t_{0K}) \\
&= C_i\left(t_0^-\right) + \frac{1}{2\pi i \hbar v_0} \tilde{M}_{i,j}\left(\frac{\omega_{i,j}}{v_0}\right) C_j(t_{0K}) \sum_{K=1}^{N} e^{i\omega_{i,j}t_{0K}} F_{et}(\omega_{i,j})
\end{aligned} \tag{D5}$$

Where $F_{et}(\omega) = \mathfrak{F}\{f_{et}(t)\}$ is the Fourier transform of the single QEW probability function. The incremental probability amplitude in (D5) averages to zero for random $t_{0K}$, except when:

$$\omega_{i,j} = n\omega_b \tag{D6}$$

where

$$t_{0K} = K 2\pi / \omega_b \tag{D7}$$

namely, when the QEWs arrive to the interaction point at a rate that is a sub-harmonic of the transition frequency $\omega_{i,j}$. Take i=2, j=1 (upper and lower levels respectively), then with the approximation of small change in the amplitude:

$$C_1(t_{0K}) \cong C_1(t_0^-) = \text{const} = 1, \quad |C_2(t_0^-)| \ll 1 \tag{D8}$$

$$\left. C_2\left(t_{0N}^+\right) \right|_{\omega_{i,j} = m\omega_L} \approx N \frac{1}{2\pi i \hbar v} \tilde{M}_{2,1}\left(\frac{\omega_{2,1}}{v}\right) F_{et}(\omega_{2,1}) \tag{D9}$$

For a Gaussian QEW,

$$f_{et} = \frac{1}{(2\pi\sigma_{et}^2)^{1/2}} e^{-t^2/2\sigma_{et}^2} \tag{D10}$$

$$P_2 = N^2 \left\{ \frac{1}{2\pi \hbar v_0} \left| \tilde{M}_{2,1}\left(\frac{\omega_{2,1}}{v_0}\right) \right| \right\}^2 e^{-\omega_{i,j}^2 \sigma_{et}^2} \tag{D11}$$

**Appendix E: Numerical computation methods**

In order to check the validity of the analytical approximations, two kinds of numerical computation codes were developed for solving the FEBERI problem of interaction between a single finite size QEW and a TLS at any initial superposition state. Extension to computation of FEBERI with modulated QEWs and with multiple correlated QEWs will be reported elsewhere.

A. Solving the differential equation derived by Schrodinger equation

Integration of the differential equation of the entangled free-bound electrons state, here we solve numerically the integro-differential equation (28) for the $C'_{ip}(t)$ that was obtain by projecting Shrödinger equation to momentum states. The equation (28) is written as

$$\dot{c}_{i,p'}(t) = \frac{1}{2\pi i \hbar^2} \int dp\, \widetilde{M}_{i,j}(p'-p) c_{j,p}(t)\, e^{-i(E_{p'}-E_p-E_{ij})t/\hbar}. \tag{E1}$$

The concept is to discretize the equation and apply the Euler method, which means to calculate the equation in a finite momentum domain $(-P_{cutoff}, P_{cutoff})$ with N points sampling. The equation becomes

$$\dot{c}_{i,p_m} = \frac{1}{2\pi i \hbar^2} \sum_{n=1}^{N} \Delta p\, \widetilde{M}_{i,j}(p_m - p_n)\, c_{j,p_n}(t)\, e^{-i(E_{p_n}-E_{p_m}-E_{ij})t/\hbar}, \tag{E2}$$

where $p_n = \left(1 - \frac{2n}{N}\right) P_{cutoff}$ and $E_{p_n} = \varepsilon_0 + v_0(p_n - p_0) + \frac{(p_n - p_0)^2}{2\gamma^3 m}$. $c_{i,p_n}(t)$ is the probability amplitude for the entangled state $|i, p_n\rangle$. Thus, the total state can be written as a vector:

$$v_i(t) = [c_{i,p_1}(t), c_{i,p_2}(t), \ldots\ldots, c_{i,p_N}(t)]^T, \tag{E3}$$

for the initial state $c_{i,p_n}(t_0) = C_i(t_0) c_{p_n}^{(0)}$ at $t_0 = t_<$, $c_{p_n}^{(0)} = \frac{1}{(2\pi\sigma_p^2)^{1/4}} \exp\left[-\frac{(p_n-p_0)^2}{4\sigma_p^2}\right]$. Summing all the scattering process from $p_n$ to $p_m$, the differential equation can be Expressed in tensor form as:

$$\frac{d}{dt} v_i(t) = U^{ij}(t) v_j(t), \tag{E4}$$

where $U^{ij}(t)$ is the matrix describing the evolution of the state $v_j(t)$. Its matrix element is

$$U^{ij}_{nm}(t) = \frac{\Delta p}{2\pi i \hbar^2} \widetilde{M}_{ij}(p_m - p_n) e^{-i(E_{p_n}-E_{p_m}-E_{ij})t/\hbar} \tag{E5}$$

With discretization of the time domain, the evolution of entangled free-electron and bound electron system is expressed as

$$\begin{pmatrix} v_1(t_{i+1}) \\ v_2(t_{i+1}) \end{pmatrix} = \begin{pmatrix} 1 & U^{12}(t_i)\Delta t \\ U^{21}(t_i)\Delta t & 1 \end{pmatrix} \begin{pmatrix} v_1(t_i) \\ v_2(t_i) \end{pmatrix}, \tag{E6}$$

with $\Delta t = t_{i+1} - t_i$.

B. Solving Heisenberg equation for the density matrix of the entangled free-bound electrons state.

In this method, the density matrix formalism is utilized to arrive at a numerical solution for the FEBERI. This facilitates the investigation of multiple QEWs. In the density matrix formalism, the Schrödinger equation (1) in the main text is generalized to the Liouville equation

$$\frac{d\hat{\rho}_{fb}(t)}{dt} = -\frac{i}{\hbar}\left[\hat{H}, \hat{\rho}_{fb}(t)\right], \tag{E7}$$

where $\hat{H} = \hat{H}_0 + \hat{H}_I$, $\rho_{fb}(t)$ represents the combined state of free electron and bound electron. Since the state of free electron is a continuous variable state, its Hilbert space is infinite. Computer simulation requires the free electron states to be discretize and sampled. The free electron initial wave function is expressed as:

$$|\psi_f\rangle = \sum_{n=1}^{N} c_{p_n}^{(0)} |\psi_{p_n}\rangle, \tag{E8}$$

where $N$ is the truncated dimension and the normalized coefficients $c_{p_n}^{(0)}$ contributes to the Gaussian distribution of Eq. (12) in the main text, i.e. $c_p^{(0)} = \langle p_n|\psi_f\rangle$. The free Hamiltonian of the free electron is diagonal in the momentum space $\{|p_n\rangle, n = 1, \cdots, N\}$

$$H_{0F}^{(m,n)} = \langle p_m|\hat{H}_{0F}|p_n\rangle = \delta_{m,n}\left[\epsilon_0 + v_0 \cdot (p_n - p_0) + \frac{1}{2\gamma_0^3 m}(p_n - p_0)^2\right]. \tag{E9}$$

The eigenstates of the bound electron are denoted by $|1\rangle$ and $|2\rangle$. This satisfies the eigenfunction $\hat{H}_{0B}|j\rangle = E_j|j\rangle$ with $E_j = \hbar\omega_j$ being the energy of state $|j\rangle$ ($j = 1,2$). The general state of the bound electron is $|\Psi_B\rangle = \sum_{j=1}^{2} C_j |j\rangle$, where the coefficients satisfy $\sum_{j=1}^{2} |C_j|^2 = 1$. The free Hamiltonian of the bound electron in the basis $\{|1\rangle, |2\rangle\}$ is

$$\hat{H}_{0B} = \begin{pmatrix} E_1 & 0 \\ 0 & E_2 \end{pmatrix}. \tag{E10}$$

The free Hamiltonian of the combined system is $\hat{H}_0 = \hat{H}_{0F} \otimes \hat{I}_2 + \hat{I}_N \otimes \hat{H}_{0B}$ with $\hat{I}_2$ and $\hat{I}_N$ being the identity operators of dimension 2 and N, respectively.

The interaction Hamiltonian between the free electron at location $\mathbf{r}_q$ and the bound electron with a dipole $\boldsymbol{\mu}$ at location $\mathbf{r}_\mu$ is

$$\hat{H}_I = -\frac{e\hat{\mathbf{r}} \cdot \hat{\boldsymbol{\mu}}}{4\pi\epsilon_0 r^3}, \tag{E11}$$

where $\hat{\mathbf{r}} = \hat{\mathbf{r}}_\mu - \hat{\mathbf{r}}_q$ and $\hat{\boldsymbol{\mu}} = e\,\hat{\mathbf{r}}'$ with $\hat{\mathbf{r}}'$ being the dipole's length vector. For the considered model in Fig. 1 of the main text, $\hat{\mathbf{r}}_q = z\,\hat{\mathbf{e}}_z$ and $\hat{\mathbf{r}}_\mu = \hat{\mathbf{r}}_\perp$. By using the approximation $|\hat{\mathbf{r}}'| \ll |\hat{\mathbf{r}} - \hat{\mathbf{r}}'| \approx (r_\perp^2 + \gamma^2 z^2)^{1/2}$, one obtains:

$$\hat{H}_I = -\frac{e^2}{4\pi\epsilon_0} \frac{\hat{\mathbf{r}}' \cdot (\hat{\mathbf{r}}_\perp - \gamma z\hat{\mathbf{e}}_z)}{(r_\perp^2 + \gamma^2 z^2)^{3/2}} \equiv \hat{\mathbf{f}}(z) \cdot \hat{\mathbf{g}}(\mathbf{r}'). \tag{E12}$$

We defined $\hat{\mathbf{f}}(z) = \frac{e}{4\pi\epsilon_0} \frac{\hat{\mathbf{r}}_\perp - \gamma z\hat{\mathbf{e}}_z}{(r_\perp^2 + \gamma^2 z^2)^{3/2}}$ and $\hat{\mathbf{g}}(\mathbf{r}') = -e\,\hat{\mathbf{r}}'$. The matrix elements of the interaction Hamiltonian in the z and r' basis are

$$\langle z_m \otimes r'_l|\hat{H}_I|z_n \otimes r'_s\rangle = \langle z_m|\hat{\mathbf{f}}(z)|z_n\rangle \otimes \langle r'_l|\hat{\mathbf{g}}(\mathbf{r}')|r'_s\rangle = \delta_{mn}\hat{\mathbf{f}}(z) \otimes \delta_{ls}\hat{\mathbf{g}}(\mathbf{r}'). \tag{E13}$$

In order to obtain the interaction Hamiltonian in the basis $\{|p_n\rangle, n = 1, \cdots, N\} \otimes \{|1\rangle, |2\rangle\}$, we first consider the matrix elements of $\hat{\mathbf{f}}(z)$ in the momentum space

$$H_{IP}^{(m,n)} = \langle p_m|\hat{\mathbf{f}}(z)|p_n\rangle = \sum_{l,s}\langle p_m|z_l\rangle\langle z_l|\hat{\mathbf{f}}(z)|z_s\rangle\langle z_s|p_n\rangle = \sum_{l,s} \hat{F}_{ml}\langle z_l|\mathbf{f}(z)|z_s\rangle \hat{F}_{sn}^\dagger. \tag{E14}$$

Expressed in matrix form it becomes

$$\widehat{H}_{IP} = \widehat{F}\left[\delta_{mn}\hat{\mathbf{f}}(z)\right]\widehat{F}^{\dagger}, \tag{E15}$$

where $\widehat{F}$ is the Discrete Fourier Transform matrix defined as $F_{mn} = \frac{\exp(-\frac{2\pi i}{N}mn)}{\sqrt{N}}$ and satisfies $\widehat{F}^{-1} = \widehat{F}^{\dagger}$. For the longitudinally and transversely aligned dipoles, the analytical expressions of such Fourier transformation is given in Appendix A with a slight different that here we extract $-er_{i,j}$ to form the dipole matrix elements (see below). Similarly, the matrix elements of $\hat{\mathbf{g}}(\mathbf{r}')$ in the basis $\{|1\rangle, |2\rangle\}$ are

$$\begin{aligned}H_{IB}^{(i,j)} = \langle i|\hat{\mathbf{g}}(\mathbf{r}')|j\rangle &= \sum_{l,s}\langle i|r'_l\rangle\langle r'_l|\hat{\mathbf{g}}(\mathbf{r}')|r'_s\rangle\langle r'_s|j\rangle \\ &= -e\sum_{l,s}\langle i|r'_l\rangle\delta_{ls}\hat{\mathbf{r}}'_l\langle r'_s|j\rangle = -e\sum_{l}\langle i|r'_l\rangle\hat{\mathbf{r}}'_l\langle r'_l|j\rangle.\end{aligned} \tag{E16}$$

If we replace the sum by integral, then $H_{IB}^{(i,j)} = -er_{i,j} = -e\int \varphi_i^*(r')r'\varphi_j(r')d^3r' \equiv \mu_{i,j}$. This is the dipole matrix elements as defined in Eq. (26), where $\varphi_i(r') = \langle i|r'_l\rangle$ ($i = 1,2$).

Now that the interaction Hamiltonian (25) in the basis $\{|p_n\rangle, n = 1, \cdots, N\} \otimes \{|1\rangle, |2\rangle\}$ can be constructed as

$$\widehat{H}_I = \widehat{H}_{IP} \otimes \widehat{H}_{IB}. \tag{E17}$$

Note that the whole Hamiltonian of the combined system is time independent. Thus, the solution of the Liouville equation (E7) is

$$\hat{\rho}_{fb}(t) = \widehat{U}(t)\hat{\rho}_{fb}(t_0)\widehat{U}^{\dagger}(t), \tag{E18}$$

where the evolution operator $\widehat{U}(t) = \exp(-\frac{i}{\hbar}\widehat{H}t)$ and the initial state of the combined system $\hat{\rho}_{fb}(t_0) = \hat{\rho}_f \otimes \hat{\rho}_b(t_0)$ with $\hat{\rho}_f = |\psi_f\rangle\langle\psi_f|$ and $\hat{\rho}_b(t_0) = |\Psi_B(t_0)\rangle\langle\Psi_B(t_0)|$. Because we are interested in the excitation of the bound electron, the state of the bound electron can be obtained by tracing out the free electron state

$$\hat{\rho}_b(t) = \text{Tr}_f\left[\hat{\rho}_{fb}(t)\right]. \tag{E19}$$

The transition probability is defined as $P_i = \langle i|\hat{\rho}_b(t)|i\rangle$ ($i = 1,2$). This definition is equivalent to $\int |c_{i,p'}|^2 dp'$, where the combined coefficients $c_{i,p'}$ is given in Eq.28 and the integration over momentum corresponds to the partial trace in Eq.E19. We can also obtain the state of the free electron by tracing out the state of bound electron

$$\hat{\rho}_f(t) = \text{Tr}_b\left[\hat{\rho}_{fb}(t)\right]. \tag{E20}$$

The energy increments of the free electron, bound electron and the combined system are defined as $\Delta E_F = \text{Tr}\left[\hat{\rho}_f(t)\widehat{H}_{0F}\right] - \text{Tr}\left[\hat{\rho}_f(t_0)\widehat{H}_{0F}\right]$, $\Delta E_B = \text{Tr}[\hat{\rho}_b(t)\widehat{H}_{0B}] - \text{Tr}[\hat{\rho}_b(t_0)\widehat{H}_{0B}]$ and $\Delta E = \text{Tr}\left[\hat{\rho}_{fb}(t)\widehat{H}\right] - \text{Tr}\left[\hat{\rho}_{fb}(t_0)\widehat{H}\right]$, respectively. The mutual energy is $\Delta E_I = \text{Tr}\left[\hat{\rho}_{fb}(t)\widehat{H}_I\right] - \text{Tr}\left[\hat{\rho}_{fb}(t_0)\widehat{H}_I\right]$.

It is straightforward to generalize the FEBERI to multiple QEWs, in which case a train of QEWs sequentially interacts with the bound electron. Assuming that the interaction time for the $n^{th}$ QEW is $t \in [t_<^{(n)}, t_>^{(n)}]$, then the evolution of the combined system is governed by the Liouville equation (E7)

$$\frac{d\hat{\rho}_{fb}^{(n)}(t)}{dt} = -\frac{i}{\hbar}\left[\hat{H}, \hat{\rho}_{fb}^{(n)}(t)\right], \tag{E21}$$

where the initial state of the combined system is $\hat{\rho}_{fb}^{(n)}(t_<^{(n)}) = \hat{\rho}_f \otimes \hat{\rho}_b^{(n-1)}(t_>^{(n-1)})$ with $\hat{\rho}_b^{(n-1)}(t_>^{(n-1)}) = \text{Tr}_f\left[\hat{\rho}_{fb}^{(n-1)}(t_>^{(n-1)})\right]$. The transition probability in the case of multiple QEWs is $P_2(t_>^{(n)}) = \langle 2|\hat{\rho}_b^{(n)}(t_>^{(n)})|2\rangle$.

Both computation methods were in good agreement. The computation results shown in the following figures were calculated with the factor code based on the matrix method (B). The results were checked at some points with code A, and found to be consistent.